\documentclass[fleqn,usenatbib]{mnras}

\usepackage[T1]{fontenc}
\usepackage{ae,aecompl}

\usepackage{graphicx}	
\usepackage{amsmath}	
\DeclareMathOperator{\sech}{sech}

\usepackage{amssymb}	
\usepackage{pdflscape}

\usepackage{txfonts} 

\usepackage{epstopdf}
\usepackage{journals}

\usepackage[none]{hyphenat}
\graphicspath{{img/}}
\DeclareGraphicsExtensions{.pdf,.png,.jpg, .jpeg}
\usepackage{pgf}
\usepackage{import}
\usepackage{adjustbox}
\usepackage{natbib}
\usepackage{xcolor} 
\usepackage{tikz}
\usetikzlibrary{positioning, calc}
\usetikzlibrary{external}
\tikzset{
  modelname/.style = {
    fill=lightgray, fill opacity=0.8, text opacity=1,
    font=\Huge
  },
  annotation/.style = {
    fill=lightgray, fill opacity=0.8, text opacity=1,
    font=\Large,
    rounded corners=3pt
  }
}
\usepackage[normalem]{ulem}
\usepackage{xspace} 

\usepackage[english]{babel} 
\usepackage{threeparttable}

\newcommand*\blu[0]{\hbox{$\text{bl}_{\text{u}}$}\xspace}
\newcommand*\blo[0]{\hbox{$\text{bl}_{\text{o}}$}\xspace}
\newcommand*\averg[1]{\ensuremath\left\langle #1\right\rangle}

\title[Face-on structure of barlenses and boxy bars: an insight from spectral dynamics]{Face-on structure of barlenses and boxy bars: an insight from spectral dynamics}
\author[Anton A. Smirnov, Iliya Tikhonenko, and Natalia Ya. Sotnikova]
{Anton A. Smirnov$^{1,2}$\thanks{E-mail: n.sotnikova@spbu.ru}, Iliya S. Tikhonenko$^{1}$, and Natalia Ya. Sotnikova$^{1}$\\
$^{1}$St. Petersburg State University,
Universitetskij pr.~28, 198504 St. Petersburg, Stary Peterhof, Russia\\
$^{2}$Central (Pulkovo) Astronomical Observatory of RAS, Pulkovskoye Chaussee 65/1, 196140 St. Petersburg, Russia\\
}

\date{Accepted XXX. Received YYY; in original form ZZZ}

\pubyear{2020}

\begin{document}
\label{firstpage}
\pagerange{\pageref{firstpage}--\pageref{lastpage}}
\maketitle

\setlength{\textwidth}{504.0pt}
\setlength{\columnwidth}{240.0pt}

\begin{abstract}
Based on the spectral analysis of individual orbits of stars from different $N$-body models, we show that the face-on morphology of the so-called `face-on peanut' bars (boxy bars) and barlenses is supported by different types of orbits. For `face-on peanut' bars, the so-called boxy orbits come to the fore, and they are responsible for the unusual morphology of the bar in the central regions. In the models with compact bulges, the bars show a barlens morphology in their central parts. We found that the barlens supporting orbits come in two types, one of which gives a square-like shape and the other have a rosette-like shape in the frame co-rotating with the bar. 
\textcolor{black}{Such a shape is typical for orbits around \textcolor{black}{stable
loop orbits} in nearly axisymmetric potentials only slightly distorted by the bar.}
 They were already known from some of the previous studies but their role in barlens shaping was barely investigated. Although quite simple, the rosette-like orbits are found to be the main building block of a barlens in our models. The detailed analysis of all bar orbits also allowed us to distinguish \textcolor{black}{the} x2 orbital family and isolate the structure supported by \textcolor{black}{orbits trapped around this family}. The x2 family is well-known, but, apparently, for the first time \textcolor{black}{in $N$-body models we have revealed the structure it supports by means of spectral dynamics and highlight its contribution to the barlens.} We found that \textcolor{black}{the} x2 family population increases with an increase in central matter concentration.
\end{abstract}


\section{Introduction}

Optical and near-infrared surveys indicate that from 45\% up to 80\% of disc galaxies in the local universe possess bars (e.g., \citealp{Eskridge_etal2000,Menendez-Delmestre_etal2007,Marinova_Jogee2007,Barazza_etal2008,Aguerri_etal2009}). 
A prevalence of barred galaxies should not be surprising. Usually considered $N$-body models of stellar discs are almost always unstable with respect to the bar formation, and very special conditions are needed to suppress the instability (some possibilities are discussed, for example, in \citealp{Sellwood_etal2019}). 
\par
Various numerical simulations have shown that after its formation, the bar grows in the vertical direction, thickens \citep{Combes_Sanders1981,Raha_etal1991} and takes a `boxy' or `peanut'-like (B/P) shape when viewed edge-on. The same B/P shape (B/P bulges) is also found in observations of edge-on disc galaxies (e.g., \citealp{Lutticke_etal2000}). Some of these bulges are accompanied by a pronounced X-structure (see, for example, SDSS image of NGC 128). From the observational point of view the connection between B/P bulges in edge-on galaxies and bars seen near face-on is a subject of a long discussion although it is more or less established to the present day. It is believed that B/P bulges are the vertically thickened inner parts of bars. 
It is also important to note that at intermediate orientations and even in the face-on view, the B/P bulges (seen as bars) can still have boxy or `peanut'-like shape and in this case, their structure can be directly associated with the structure of the 3D bar~\citep{Erwin_Debattista2013,Erwin_Debattista2017,Laurikainen_Salo2017,Li_etal2017}. 
It is worth noting that B/P-shaped bulges prefer to settle in early-type galaxies~\citep{Erwin_Debattista2017,Li_etal2017}. A lot of them are found among \textcolor{black}{early-type}
galaxies. Their fraction decreases significantly for later-type galaxies, they are almost never found in Scd galaxies and even later type galaxies \citep{Erwin_Debattista2017,Li_etal2017}. 
\par
Bars observed in nearly face-on galaxies do not necessarily show boxy or `peanut'-like isophotes or even X-shaped structures as, for example, IC~5240 \citep{Laurikainen_Salo2017}. Often a face-on bar demonstrates a barlens, that is, it has a barlens-like morphology in central regions with round or oval isophotes, and the barlens itself is embedded in \textcolor{black}{the main} elongated and narrow bar. Barlenses were massively introduced as a separate galactic structure only recently 
by~\cite{Laurikainen_etal2011},
although individual galaxies with barlenses have been studied for a long time (NGC~1097, NGC~4736, NGC~5728 \citealp{Shaw_etal1993}; NGC~4442 \citealp{Bettoni_Galletta1994}). 
\par
\par
Statistics of B/P-shaped bars and barlenses are available in only a few works, and they are not \textcolor{black}{commonly accepted} and can be revised for larger samples. 
In addition, the fractions of \textcolor{black}{B/P-shaped bars and bars with barlenses} are highly dependent on the galaxy inclination. For example, in the CGS sample by \citet{Li_etal2017}, among 264 galaxies with bars identified in $I$-band images, the fraction of bars flagged as buckled (B/P-shaped bars and bars with barlenses) is 40\%, regardless of the inclination. 
At the same time, the share of barlenses among barred galaxies at low and moderate inclinations is higher that the share of bars with boxy isophotes \textcolor{black}{in central areas of the galaxies} and X-structures. The fraction of barlenses substantially diminishes at high inclinations, while the fraction of galaxies with B/P/X-shaped features increases on the contrary.
\citet{Laurikainen_Salo2017} believe that such a complementarity between galaxies with B/P/X-shaped features and galaxies with barlenses indicates that these structures are the manifestation of the same phenomenon, i.e. B/P bulges and barlenses are the same structures but observed from a different viewing angle. At high galaxy inclinations, these structures are observed as B/P/X-shaped features, and viewed face-on, they are observed as barlenses. However, there exist galaxies at a low inclination, whose bars look like `face-on peanut' rather than a barlens embedded in a bar  (e.g. IC~4237, IC~5240, NGC~3227, NGC~4123, NGC~4725, \citealp{Laurikainen_Salo2017}). Thus, in early-type galaxies, there are at least two types of bars with different face-on morphology: `face-on peanuts' with X-shaped features in central regions (X) and barlenses (BL). \textcolor{black}{We should also note some important things concerning the bar morphology and the observational data. The problem is that one can consider the morphology of the barred galaxy as a whole or the morphology of the bar itself. These two are different things and the later type is obtained after the subtraction of some additional components which are not genetically associated with the bar itself (for example, a classical bulge). Whether or not the morphology of these two images is significantly different for each particular galaxy depends on the properties of these additional components. Hereinafter, we consider only the bar morphology neglecting the influence of other components.}
\par 
Currently, there are works in which the difference in the morphology of the bars in a face-on view (a `face-on peanut' or a barlens) is associated with the difference in the parameters of the underlying galaxy. 
For example, \citet{Salo_Laurikainen2017,Laurikainen_Salo2017} concluded that the introduction of a moderate bulge in the $N$-body model drastically changes the bar morphology. Models by \citet{Salo_Laurikainen2017}, which have a small classical bulge and a steep inner rotation curve slope, give BLs, whereas bulgeless models with a shallow rotation curve end up with an X-shaped feature visible even in a face-on view. 
It was enough to have a bulge-to-disc (B/D) mass ratio about 0.08 at the beginning of the simulation, so that the growing  B/P/X bulge takes a typical BL shape viewed face-on.
\par
A BL morphology can be produced in $N$-body simulations, which include gas (SPH), without the need to add any spheroidal bulge components in the initial models \citep{Athanassoula_etal2013,Athanassoula_etal2015}. Apparently, the reason for BL morphology to appear in simulations by \citet{Athanassoula_etal2013} is the increase in the gas concentration in the central area via the bar-induced inflow. However, it should be remembered that the well-developed barlenses are clearly more common in early-type S0s, which have consumed most of their gas but have classical bulges \citep{Laurikainen_Salo2017}. 
\par
\par
\par
The observational data indicate that barlenses come in different types themselves. For example, \citet{Laurikainen_Salo2017} give examples of nearly face-on galaxies with barlenses and a weak X-shaped feature in the unsharp-masked images. These are intermediate cases between X and BL.  
\par
\par
The existence of the connection between the potential of the parent galaxy and the face-on bar morphology is a direct consequence of the fact that in different potentials different families of orbits will be stable. 
Although the whole picture is also complicated by the fact that different stable orbital families have different populations. There is a large body of literature on orbits supporting a \textcolor{black}{2D} or 3D bar. Different studies in the field of nonlinear dynamics (for example, \citealp{Contopoulos_Papayannopoulos1980,Athanassoula_etal1983,Skokos_etal2002a,Skokos_etal2002b,Patsis_etal2002,  Kaufmann_Patsis2005, Patsis_Harsoula2018,Patsis_Athanassoula2019}) have identified a lot of possible families of periodic and quasi-periodic orbits with a certain morphology in the bar potential while $N$-body simulations, especially those where the orbital frequencies of bodies are determined explicitly (for example, \citealp{Voglis_etal2007,Harsoula_Kalapotharakos2009,Contopoulos_Harsoula2013,Portail_etal2015b,Gajda_etal2016,Valluri_etal2016,Chaves-Velasquez_etal2017,Abbott_etal2017,Lokas2019}), have revealed what is realised in conditions close to that in real galaxies. A comparative analysis of the orbital composition of various models, as in \citet{Parul_etal2020}, is a key point for understanding how the different types of bar morphology can be connected with the physical properties of real galaxies. And if the features of the vertical structure are determined by the difference in the distribution function of the orbits over the ratio of the vertical oscillations frequency to the in-plane frequency \citep{Parul_etal2020}, then the features of the face-on morphology must be determined by the distribution function of the in-plane frequency ratios.
\par
We analyse the orbital composition of four $N$-body models, two from \citet{Smirnov_Sotnikova2018} and two new models. \textcolor{black}{These models have different bar morphologies --- from a face-on peanut-shaped bar to a pronounced barlens embedded in a narrow bar.} Using dominant frequencies we identified all orbital groups which do not enter to the outer disc. For all these orbits we \textcolor{black}{construct} the distributions over the ratios of the in-plane frequencies and show how the dominance of one or another orbital group determine the face-on morphology of a bar and its features.
\par
In Section~\ref{sec:methods} we present our numerical models and the overall picture of their evolution. 
Section~\ref{sec:freq_analysis} contains the description  
of the frequency analysis method \textcolor{black}{together with various technical details of its implementation.} In Section~\ref{sec:bar_ring_disc} we give the details of our algorithm for identifying \textcolor{black}{particles-``stars'' captured by the bar and dissect each numerical model onto the bar component assembled by such particles and the rest.} 
In Section~\ref{sec:plane_res} we \textcolor{black}{perform a comparative analysis of various numerical models in terms of their} bar face-on morphology and the distributions of frequency ratios of all orbits contributing to the bar component and identify all components in a bar, which are supported by different types of orbits. 
In Section~\ref{sec:bar_anatomy} and Section~\ref{sec:barlens_anatomy}, based on the example of one particular numerical model we \textcolor{black}{establish a straightforward connection between}
the orbital \textcolor{black}{content of the bar} and \textcolor{black}{the various substructures of the bar:} an elongated bar, a box-shaped bar, and a barlens embedded into the bar.
In Section~\ref{sec:comparison} we summarise the trends in the dominance of different orbital groups \textcolor{black}{depending on the particular numerical model} and show how the contribution of these groups to the overall bar changes from model to model. 
In Section~\ref{sec:orbits}, we discuss \textcolor{black}{how the orbital groups distinguished in the present work are related to the well-known families of orbits distinguished in the previous orbital studies of the bar} as well as the morphological features of a bar produced by such orbits. Also, the typical orbits that can constitute a barlens are presented in this Section.
In Section~\ref{sec:discussion}, we discuss the nature of the central barlens from the point of view of the orbits, which are gradually involved in this structure, and give an interpretation of some of the observed features of barlenses. 
Finally, in Section~\ref{sec:conclusions} we summarise our results.

\section{Methods}
\label{sec:methods}

\subsection{Numerical models}
\begin{figure}
\begin{minipage}{.48\textwidth}
\centering
\includegraphics[width=\textwidth]{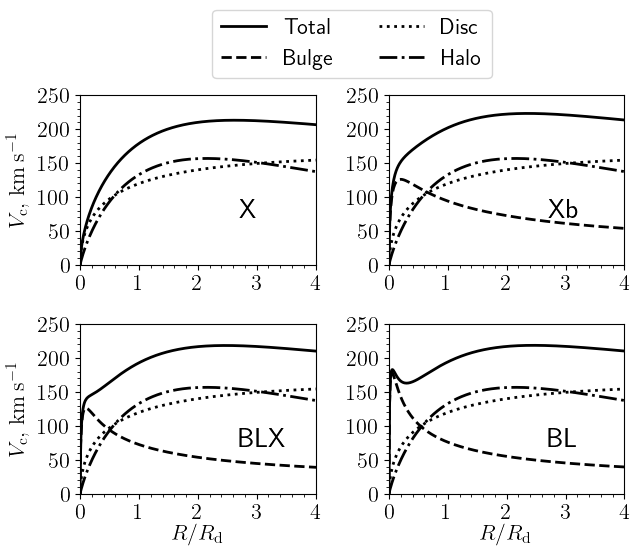}\\
\end{minipage}
\caption{Rotation curves of the numerical models used in the present study. Different line styles are used to highlight the contribution of individual components. Models are arranged by the increase in the steepness of their rotation curves. Models notation (``X'', ``Xb'', etc.) follows that of the main text.}
\label{fig:vel_curve}
\end{figure}

To understand the physical structure of barlenses we first need to obtain appropriate models where barlenses arise. Here we rely on the results of the previous numerical and observational studies. According to~\citet{Salo_Laurikainen2017} one of the crucial parameters determining the \textcolor{black}{formation} of barlenses is the steepness of the \textcolor{black}{initial} rotation curve \textcolor{black}{which implies the presence of the central matter concentration (CMC)}. \textcolor{black}{The central matter concentration can be caused by the gas that has lost its angular momentum and has been gathered in the centre, \textcolor{black}{or} by a dark halo with a cuspy density distribution, and \textcolor{black}{or} by a compact classical bulge. The study of~\citet{Salo_Laurikainen2017} focused on the last option and the connection between barlenses and compact classical bulges.} Naturally, \textcolor{black}{the compactness of a classical bulge} depends on the bulge contribution ($B/T$) as well as on how much its mass is actually compressed ($r_\mathrm{b}/h$). 
We follow~\cite{Salo_Laurikainen2017} in the present work and limit ourselves to the pure stellar models with dark halo profiles of NFW-type~\citep{NFW}. 
In this way, we try to ensure that barlenses we study have the same physical nature, the same formation mechanism and are supported by the same types of orbits (if any) as in~\cite{Salo_Laurikainen2017}. This is important because the models considered by~\cite{Salo_Laurikainen2017} gave barlenses similar to those observed in real galaxies and were already compared with observational data~\citep{Salo_Laurikainen2017,Laurikainen_etal2018}.
\par
\cite{Salo_Laurikainen2017} chose the exact values of classical bulges parameters based on the results of the 2D decomposition of S4G galaxies from their previous works~\citep{Laurikainen_etal2014,Salo_etal15}. They considered two types of bulges, one with \textcolor{black}{a bulge-to-disc mass ratio} $B/D=0.08$ (\textcolor{black}{$B/T\approx0.074$}) and the other with $B/D=0.01$ (\textcolor{black}{$B/T\approx0.01$}) (both bulges have \textcolor{black}{a bulge effective radius $r_\mathrm{B}/h_\mathrm{D}=0.07$, where $h_\mathrm{D}$ denotes the scale length of the disc}). These values seem rather low for a typical classical bulge (see, for example, \citealt{Gao_etal2020}). However,~\cite{Laurikainen_etal2018} argued that $B/T$ values strongly depend on the applied photometric model. More precisely, $B/T$ strongly depend on whether or not barlens is included in the photometric model as a separate component. An overestimated B/T value can be two or three times greater (or even greater) than it actually is if the barlens is not taken into account.~\cite{Erwin_etal2015} also obtained a similar result in case of composite bulges (when a galaxy posses a pseudo bulge and a classical bulge at the same time). Therefore, values of $B/T$ in the range from about $0.01$ and up to $0.1$ seem reasonable from the perspective of the refined 2D decomposition.
\par
In this work, we consider four different numerical models, varying the parameters of classical bulges while all other components being the same. \textcolor{black}{These models were used as initial configurations for the self-consistent evolution that will be described in the next subsection.} We found by trial and error that these four cases
demonstrate the fairly gradual transition of the bar morphology. This is important because it will allow us to obtain a general picture of how and why barlenses appear in various numerical models. \par
The details of \textcolor{black}{our initial model building}
are the following. We took as basis two models from our previous work~\citep{Smirnov_Sotnikova2018}. These models consisted of an exponential disc isothermal in the vertical direction,
\begin{equation}
\rho_\mathrm{d}(R,z) = \frac{M_\mathrm{d}}{4\pi R_\mathrm{d}^2 z_\mathrm{d}} \cdot \exp(-R/R_\mathrm{d}) \cdot \sech^2(z/z_\mathrm{d}) \,,
\label{eq:rho_disk} 
\end{equation}
where $M_\mathrm{d}$ is the total mass of the disc and $R_\mathrm{d}$ and $z_\mathrm{d}$ are scale lengths in radial and vertical directions, respectively. The dark halo was modelled by a truncated sphere with the density profile close to the NFW profile~\citep{NFW} but with a slightly steeper inner slope,
\begin{equation}
\rho_\mathrm{h}(r) = 
\frac{C_\mathrm{h}}
{(r/r_\mathrm{s})^{\gamma_0}
\left((r/r_\mathrm{s})^{\eta}+1\right)^
{(\gamma_{\infty}-\gamma_0)/\eta}} \,,
\label{eq:NFW}
\end{equation}
where $r_\mathrm{s}$ is the halo scale radius, $\eta$ is the halo transition exponent, $\gamma_0$ is the halo inner logarithmic density slope, $\gamma_{\infty}$ is the halo outer logarithmic density slope, $C_\mathrm{h}$ is the parameter defining the total mass of the halo $M_\mathrm{h}$. We adopted the following values: $\eta=4/9$, $\gamma_0=7/9$, $\gamma_{\infty}=31/9$. $C_h$ was chosen to produce a reasonable dark halo profile with $M_\mathrm{h}(r < 4R_\mathrm{d})/M_\mathrm{d} \approx 1.5$ (see Fig.~\ref{fig:vel_curve}).
\par 
One model was pure bulgeless (hereinafter model X) while the second one possessed a classical bulge (model Xb) of a \cite{Hernquist1990} profile,
\begin{equation}
\rho_\mathrm{b}(r) = \frac{M_\mathrm{b}\, r_\mathrm{b}}{2\pi\,r\,(r_\mathrm{b} + r)^3} \,,
\end{equation} 
where $r_\mathrm{b}$ is the scale parameter and $M_\mathrm{b}$ is the total bulge mass. $B/T$ value was rather large for this model, $M_\mathrm{b}/M_\mathrm{d}=0.2$, compared to typical values from~\cite{Laurikainen_etal2018}. But the bulge scale length was also rather large, $r_\mathrm{b}/R_\mathrm{d}=0.2$, so the bulge mass was dispersed across rather large volume. 
The velocity curve \textcolor{black}{in this model is steeper than in the bulgeless model X} (see Fig.~\ref{fig:vel_curve}). 
\par
Two new models we construct for this study were the same type but with more concentrated bugles. The first one has $M_\mathrm{b}/M_\mathrm{d}=0.1$, $r_\mathrm{b}/R_\mathrm{d}=0.1$ (BLX model) and the second one has $M_\mathrm{b}/M_\mathrm{d}=0.1$, $r_\mathrm{b}/R_\mathrm{d}=0.05$ (BL model), respectively. According to the CMC, the models can be arranged in the following way: BL > BLX > Xb > X (see Fig.~\ref{fig:vel_curve}).
\par
We consider dynamically cold discs with Toomre parameter value $Q=1.2$ at $R=2R_\mathrm{d}$. The radial velocity dispersion profile was chosen to obey an exponential law:
\begin{equation}
\sigma_R=\sigma_0 \exp(-R/2R_\mathrm{d}),
\end{equation}
where $\sigma_0$ is the dispersion of radial velocities in the centre. Its value is derived from the condition on the Toomre parameter, $Q(R=2R_\mathrm{d})=1.2$.


\subsection{Simulations}

\begin{figure}
\begin{minipage}{.48\textwidth}
\centering
\includegraphics[width=\textwidth]{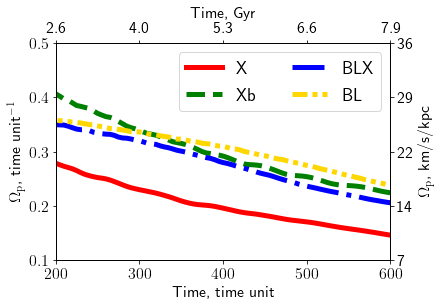}\\
\end{minipage}
\caption{Evolution of the bar pattern speed in different models in simulation units (\textit{left axis}) and in usual km/s/kpc (\textit{right axis}).}
\label{fig:pattern_speed}
\end{figure}
\begin{figure*}
\centering
\includegraphics{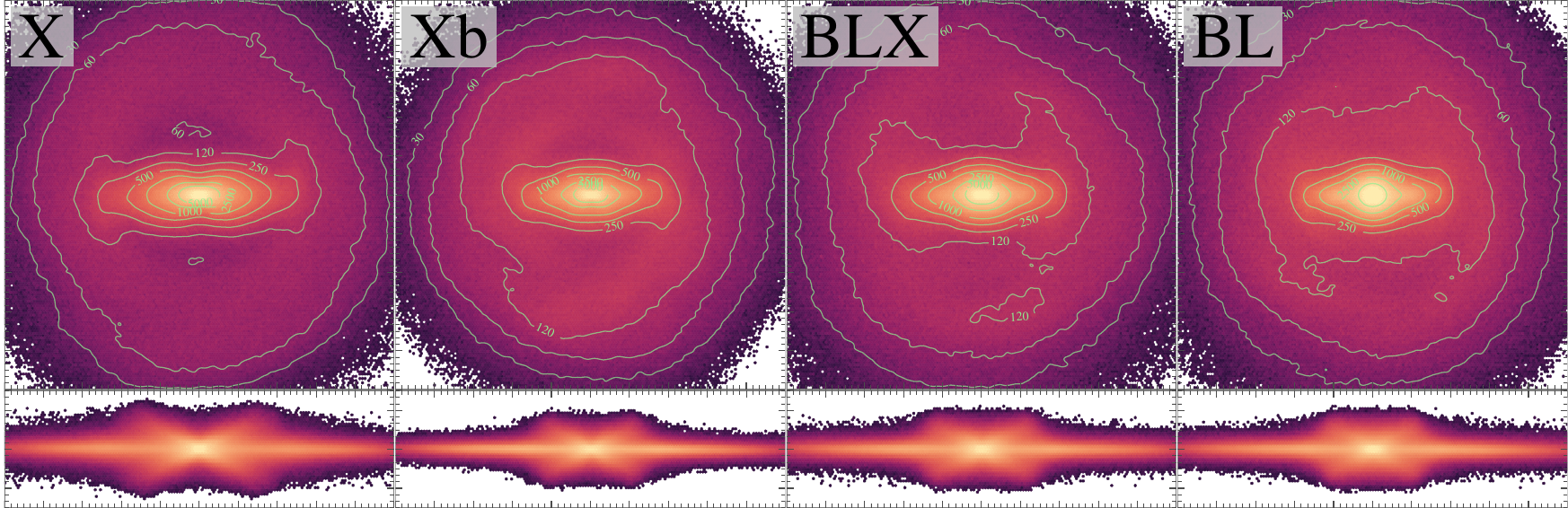}
\caption{Face-on views of different numerical models at $t=450$ (6~Gyr) displayed in the square $(xy)=(-5,5)\times(-5,5)$,
side-on views in $(xz) = (-5,5)\times(-1.5, 1.5)$. }
\label{fig:discsoverview}
\end{figure*}

Here we briefly describe various aspects of the simulations. The whole procedure is mostly repeating that from~\cite{Smirnov_Sotnikova2018} and we refer an interested reader to this work. An $N$-body representation of each model was prepared via {\tt{mkgalaxy}} code of~\cite{McMillan_Dehnen2007}. This code is a part of {\tt{NEMO}} project~\citep{Teuben_1995} and free to use. We use 4$kk$ particles for the disc and 4.5$kk$ for the halo. Bulge particles have the same mass as disc particles. The total number of bulge particles is then determined according to $M_\mathrm{b}/M_\mathrm{d}$ ratio. After self-consistent $N$-body representation of each model was obtained the evolution of the models was calculated via {\tt{gyrfalcON}} code~\citep{Dehnen2002}. We
use an adaptive time step with the maximal allowed value equal to 0.125 in simulation units which translates into $\approx$ 1.7 Myr if we assume $R_\mathrm{d}=3.5$ kpc and 
$M_\mathrm{d}=5 \cdot 10^{10} M_\mathrm{\sun}$. Hereinafter we measure time intervals in time units (t.~u. for short) of our simulations, $1$~t.~u.${\approx}$14 Myr, and distances in units of $R_\mathrm{d}$, $1$~length unit$=3.5$ kpc. Length unit is also denoted by l.~u. for short. Here and below, if a variable goes without a unit of measurement, then its value is measured in the corresponding units of simulation. The softening length values for the disc, $\varepsilon_\mathrm{d}$, and the halo, $\varepsilon_\mathrm{h}$, were scaled to our number of particles from the values given in~\cite{McMillan_Dehnen2007} for the same type of models. The resulted values were about $3.7 \cdot 10^{-3}R_\mathrm{d}$ or $\approx$13 pc for the disc and $12.9 \cdot 10^{-3}R_\mathrm{d}$ or $\approx$45 pc for the halo, respectively. We note that the choice of softening length is important for our problem because \cite{Salo_Laurikainen2017} found that face-on bar morphology is actually dependent on it. They find that $\varepsilon_\mathrm{d}=0.02R_d$ do not lead to a barlens while the smaller values indeed give it for their model. Our values of $\varepsilon$ are considerably smaller than $0.02$ and therefore, there should be no problems with barlens manifestation due to insufficient softening length. 
The evolution of the models was followed up to 8 Gyr. Each model is prone to a bar instability and a bar inevitably forms after several Gyr in each of them. It is important that each model leads to a different morphological type of a bar depending on the CMC (Fig~\ref{fig:discsoverview}). More concentrated models give rise to a bar with a barlens while those without classical bulges lead to a peanut-shaped bar (in a face-on view).
\par
To study the orbital composition of the bar we apply the methods of spectral dynamics~\citep{Binney_Spergel1982} in the following subsection. One of the main concerns of such an analysis is how reliable the frequencies we obtain in case of evolving bar/disc. The orbital frequencies of the bar particles are tied to the bar pattern speed. If the pattern speed varies then the frequencies vary too. The general strategy to deal with this problem is to choose some time interval where the bar pattern speed is more or less established. Fig~\ref{fig:pattern_speed} show how the pattern speed of the bar varies in our models. One can see that \textcolor{black}{it decreases by about half in 6~Gyr for different models} which implies some frequency shift. However, in our previous work~\citep{Parul_etal2020}, we found that the actual frequency shift is small for most of the particles and its value is about the frequency measurement error (see figure~5 from~\citealt{Parul_etal2020}). In \cite{Parul_etal2020} the frequency shifts were estimated for the time interval $t=400-500$ t.u. For our convenience, we select this time interval to apply the frequency analysis in the present work.


\section{Analysis of dominant frequencies}
\label{sec:freq_analysis}
\par
The analysis we carry out is almost the same as in~\cite{Parul_etal2020}. The key features are the following: 

\begin{enumerate}
\item We work with self-consistent $N$-body snapshots (\textbf{not} the frozen potential).  
\item All 4$kk$ particles composing the disc are processed, nothing excluded. 
\item Each orbit is characterised by four time series of $x,y,z$ and $R$ (cylindrical radius) coordinates. Each orbit then can be characterised by the set of the dominant frequencies, $f_x, f_y, f_z$ and $f_\mathrm{R}$, respectively.
\item We work in the reference frame co-rotating with the bar. It simplifies an analysis since we do not need to modify the obtained frequencies according to the changing pattern speed of the bar. Plus $2\pi f_x \approx \Omega - \Omega_\mathrm{p}$ for the most of bar orbits~\citep{Gajda_etal2016} which helps with the interpretation of the results.
\end{enumerate}
The time series for each coordinate consisted of 801 data points, from $400$ to $500$ time units or from $\approx$5.3 Gyr to $\approx$6.6 Gyr. The orbit spectra were obtained by means of Fast Fourier Transform (FFT). The Nyquist frequency value $\omega_c$ was about $1800$ km/s/kpc while the frequency grid step $\Delta \omega$ was about $4.5$ km/s/kpc. The frequency resolution (due to discrete frequency grid) was improved using the procedure similar to the zero-padding but less time consuming (see~\citealt{Parul_etal2020} for the details). The resulting frequency resolution $\Delta \omega_\mathrm{imp}$ was about $0.45$ km/s/kpc. 
After we obtained the spectral characteristics of each orbit the next goal is to accurately clear bar particles from disc/ring debris. This will be done in the next section. 
\par
As we see below, for some important orbits we also need to distinguish the spectral line, which is second in amplitude. There is nothing strange for an orbit to have several spectral lines (see~\citealt{Binney_Spergel1982}) but this point is rather ignored in some recent studies, unfortunately. 
We extract the frequency of the second line in a simple way, subtracting the contribution of the dominant wave from the original spectrum and finding the maximum of the residue spectrum.


\section{Bar, ring, and outer disc} 
\label{sec:bar_ring_disc}
Before we start to analyse the orbits composing the 3D bar, we need to separate the bar from the disc.
We will describe this procedure in the current section.
\par
\begin{figure}
\centering
\includegraphics{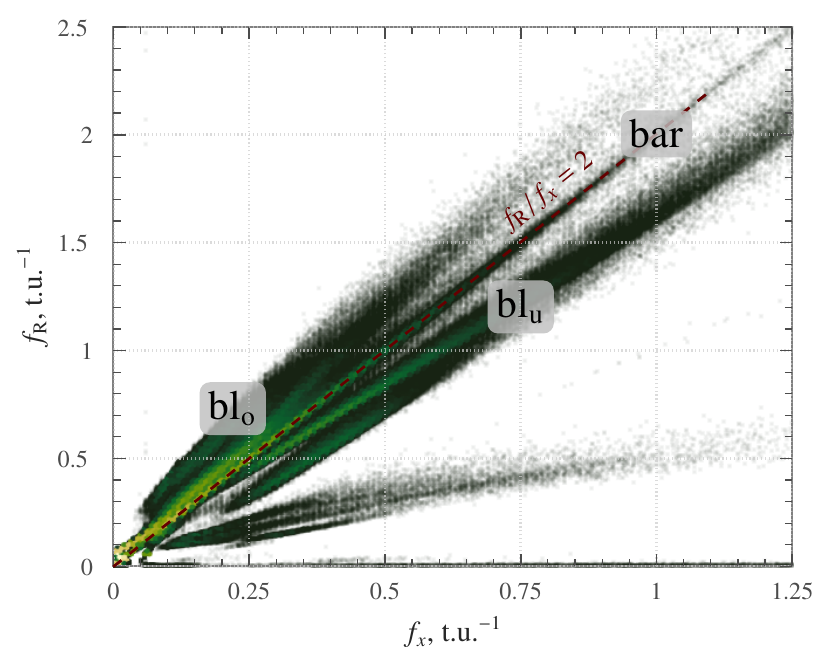}
\caption{2D distribution of orbits over $f_\mathrm{R}$ and $f_x$ frequencies for the model with the barlens (model BL). The dark red dashed line corresponds to the perfect bar frequency ratio $f_\mathrm{R}:f_x=2:1$. \textcolor{black}{The two orbital groups, which are located along the thick rays above and below the dark red dashed line, are denoted as \blo and \blu. These groups 
will be described in detail in Section~\ref{sec:main_groups_bar}.}}
\label{fig:fRx_all}
\end{figure}

\citet{Portail_etal2015b,Gajda_etal2016} have identified the bar in their $N$-body simulations by the condition $f_\mathrm{R}/f_x = 2.0 \pm 0.1$. Fig.~\ref{fig:fRx_all} shows 2D distribution of frequencies $f_\mathrm{R}$ and $f_x$ for our model BL. A bright straight line corresponding to the $f_\mathrm{R}/f_x\approx2.0$ is clearly visible. However, the situation with the identification of the bar, when it comes to all the particles in the model, is more complicated. First, particles in the region $f_\mathrm{R} \approx 0$ and $f_x \approx 0$ do not belong to the bar, but to the outer regions of the disc. Secondly, above and below the resonance line $f_\mathrm{R}/f_x = 2.0$, there is a significant number of particles that are located in the central areas of the bar, and they cannot be neglected while studying orbits that support the complex structure of the bar. 
\par
\begin{figure}
\includegraphics{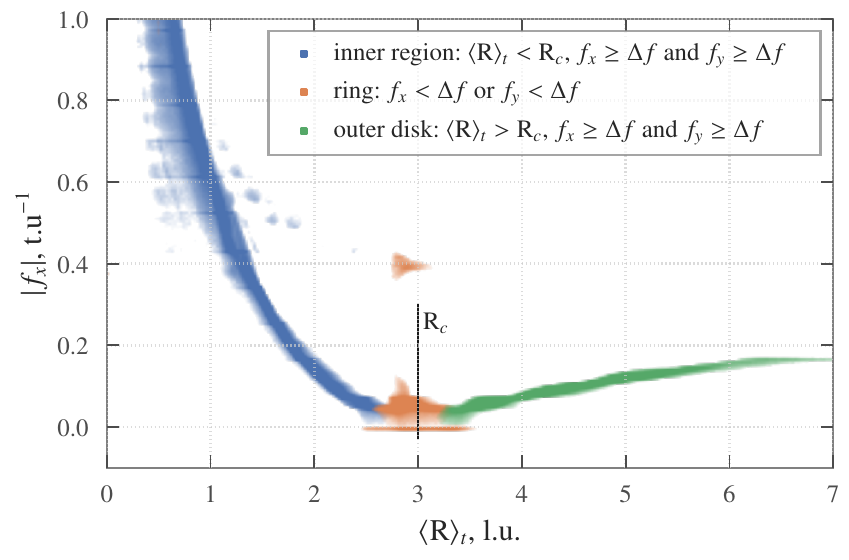}
\caption{2D distribution of all disc particles on $\averg{\mathrm{R}}$-$f_x$ plane for the model BL, showing the major subsets: the ring (in orange colour), the outer disc (green) and the inner part (blue). $\mathrm{R}_c$ corresponds to the co-rotation radius adopted for this model. $\Delta f$ corresponds to the finite frequency resolution (see the text for details).}
\label{fig:regionsplit}
\end{figure}

Thirdly, due to our choice of the reference frame co-rotating with the bar, the model has a specific radius, at which the angular frequency of the frame matches the angular frequency of the particles in the disc~---
the radius of co-rotation (CR hereinafter). This region \textcolor{black}{has a shape of a broad ring, sitiated between} the bar \textcolor{black}{and} the outer disc \citep[Fig.9]{Ceverino_Klypin2007}. In the vicinity of CR, particles are unable to finish a turn around the centre during the considered time interval. Consequently, their orbital frequencies $f_x$ or $f_y$ tend to be smaller than the frequency resolution $\Delta f = 1/\delta t$, where $\delta t$ is the time interval on which we measure the frequencies.
Using this feature, we can distinguish three major groups of orbits in all of our models. These groups are shown by different colours in Fig.~\ref{fig:regionsplit},
where we plot $|f_x|$ against time-averaged orbital radius $\averg{R}_t$ for all 4$kk$ disc particles in the model BL.
Given all of the above, the procedure for separating the bar from the outer areas of the disc includes several steps.
\begin{enumerate}
\item The particles with $f_x < \Delta f$ or $f_y < \Delta f$ constitute a ring \textcolor{black}{(not shown here, but see figure 9 in \citealt{Ceverino_Klypin2007}) the median radius of which roughly corresponds to CR (Fig.~\ref{fig:regionsplit}).} A more precise definition of CR is not required, as the groups are well separated by average orbital radii (see Fig.~\ref{fig:regionsplit}).
Since the bars cannot extend beyond the co-rotation \citep{Contopoulos1980}, this group has been removed from further analysis.
\item We associate the particles with $f_x \geq \Delta f$ and $f_y \geq \Delta f$ that are further than CR with the outer disc.
This group does not contain any bar orbits and is irrelevant for the current study, too.
\item The group with $f_x \geq \Delta f$ and $f_y \geq \Delta f$ located inside the co-rotation radius is the main contributor to the region that contains the bar. A detailed study of orbits from this group is a primary goal of the current analysis and will be carried out in the following sections.
\end{enumerate}

\begin{figure}
\centering
\includegraphics{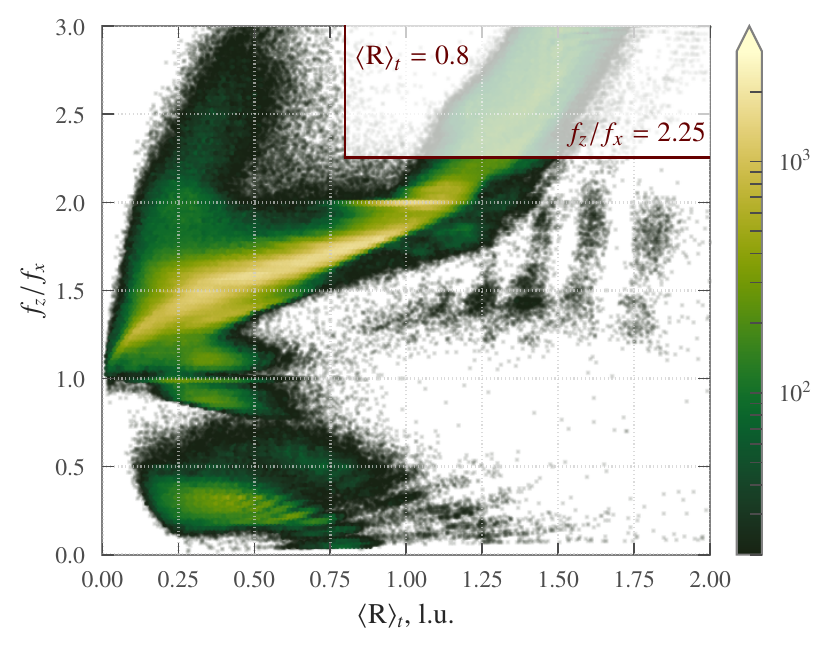}
\caption{2D distribution of particles inside CR on $\averg{\mathrm{R}}_t-f_z/f_x$ plane. 
The shaded region bounded by two red lines is excluded from further analysis.
See the main text for details.}
\label{fig:inner_2Dfzfx}
\end{figure}

We attempted to refine the outlined bar selection scheme further.
Fig.~\ref{fig:inner_2Dfzfx} shows the distribution of the orbits of the latter subset on $\averg{R}_t-f_z/f_x$ plane.
In this plot, most of the particles lie below the resonance line $f_z/f_x = 2$, which corresponds to `banana'-like orbits. The region above this line consists of two parts. The first one is a group of orbits with average radii smaller than about $0.8$ length units.
Since these orbits are located near the centre, we assume that they constitute a part of the bar and include them in further analysis.
The second one is a ``stripe'' of orbits that are further from the centre and have higher $f_z/f_x$ than `banana'-like orbits.
This group consists of orbits that inhabit the peripheral parts of the bar, where it connects to the ring.
We decided to exclude all of them using the conditions $\averg{R}_t > 0.8$ and $f_z/f_x > 2.25$ (see the shaded area in Fig.~\ref{fig:inner_2Dfzfx})
The value of the frequency ratio was chosen so as not to lose `banana'-like orbits for sure. 
This is justified because this group consists of extended orbits, while we are mainly interested in the inner regions where the barlens resides. 
Thus, for the further spectral analysis we have discarded orbits not only from the rest of the disc outside CR but also the orbits that belong to the thin part of the bar. All these orbits do not contribute to the thick B/P bulge but only to the most remote and vertically thin parts of the bar. As a result, we have left only the thick part of the bar. The $xy$ and $xz$ views for this part of the bar for all our models are shown in Fig.~\ref{fig:xy_bar}. Although the face-on structure under study has a different length along the major axis of the bar, in the vertical direction it looks equally thick and peanut-shaped for all models, and this is what is called the B/P bulge. 

\section{Face-on portrait of a bar: morphological and orbital analysis}
\label{sec:plane_res}

\begin{figure*}
\centering
\includegraphics{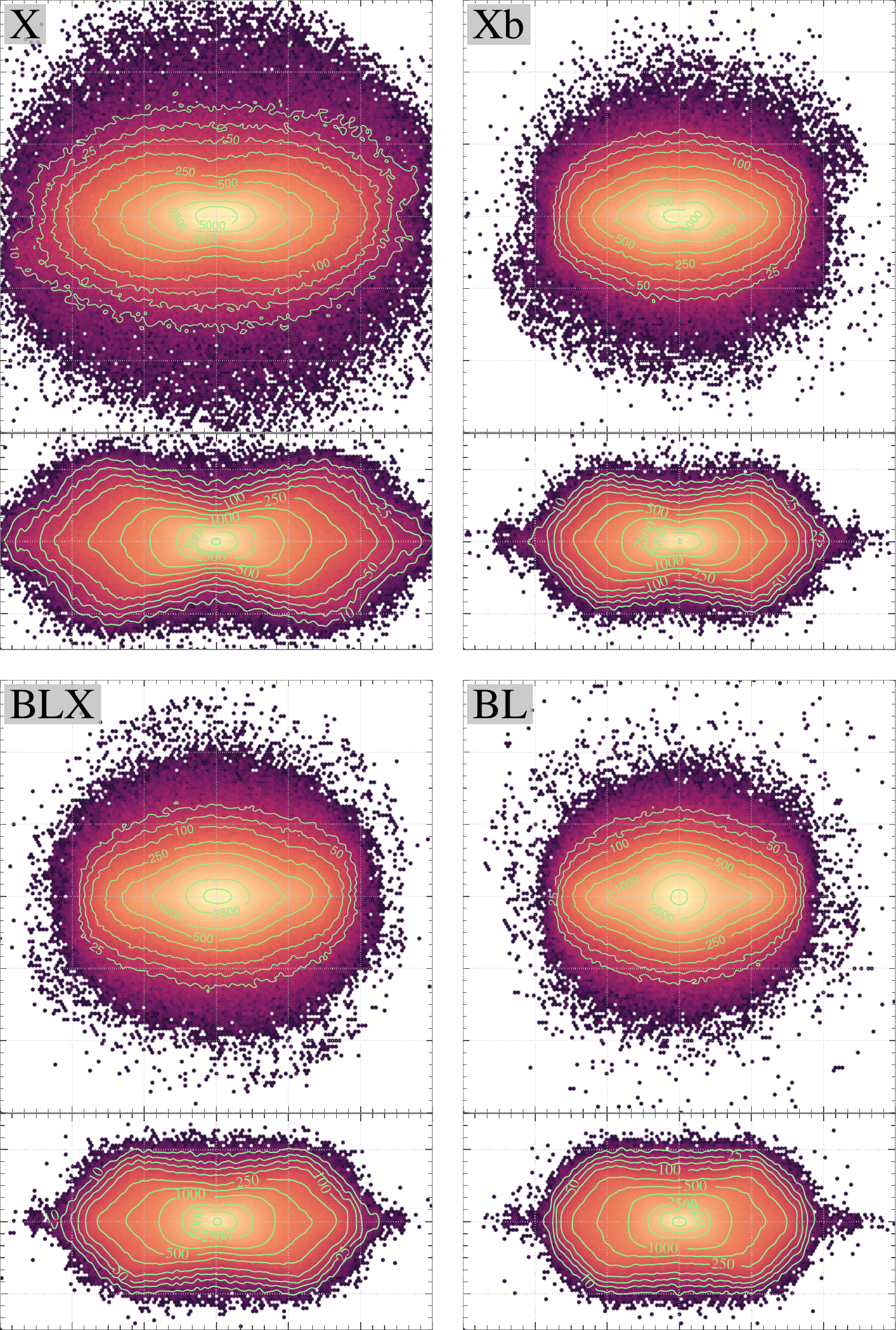}
\caption{Face-on view \textcolor{black}{of the B/P part} of a bar for X and Xb (two upper plots) and BLX and BL (two bottom plots) models at $t=450$ t.~u. (6~Gyr). All plots depict the same space area $(xy)=(-3,3)\times(-3,3)$.}
\label{fig:xy_bar}
\end{figure*}

Fig.~\ref{fig:xy_bar} shows face-on and side-on portraits of our modelled bars (the B/P part of the bars), cleaned of the ring and outer disc as well as of the particles of the flat component inside the ring, which were cut out at the initial steps of the current analysis (see Section~\ref{sec:bar_ring_disc}). All models show thick peanut- or boxy-shaped structure in the vertical direction. The side-on peanut-like shape is more typical for the X model, while the BL model with a clear face-on barlens shows box-like isophotes in the $xz$ projection, especially in the central regions. As to the face-on views, all models look different.
The model without bulge (model X) demonstrates a peanut-shaped morphology in the face-on view with an X-shaped pattern in the centre \textcolor{black}{discernible on the unsharp-masked images (not shown here due to lack of space)}. Such a morphology is mainly supported by the so-called boxy orbits (see Section~\ref{sec:orbits} and references therein). 
With an increase in B/D and a transition to a more concentrated bulge, i.e. with an increase of CMC, barlens component makes its appearance in the $xy$ view (Fig.~\ref{fig:xy_bar}, from left to right, from top to down).
\par
We expect that the observed transition in the face-on bar morphology is associated with some notable changes in the orbital composition of a bar. We understand the orbital composition as a set of orbital groups that differ in the ratios of the in-plane frequencies $f_x$, $f_y$ and $f_\mathrm{R}$. The ratio $f_\mathrm{R}/f_x$ is usually employed to separate a bar according to the condition $f_\mathrm{R}/f_x=2.0$ \citep{Portail_etal2015b,Gajda_etal2016}. Nevertheless, 
\citet{Gajda_etal2016} showed that in their models, there are additional orbits in the region of the bar that can be identified only if the frequency $f_y$ is also involved. To accurately distinguish the orbits that can be associated with a barlens, we first would like to grasp the whole picture of what orbital groups can actually constitute the bars in our models. To this aim, we want to distinguish all morphologically different orbital groups existing in the bar area for all of our models. We assume that each individual group is characterised by a unique set of frequency ratios $f_\mathrm{R}/f_x$ and $f_\mathrm{R}/f_y$. We begin with the description of a qualitative picture, that is, what types of orbits on the plane $f_\mathrm{R}/f_x$--$f_\mathrm{R}/f_y$ can constitute a bar in each of the models and what new types of orbits start to emerge depending on the model. Although the ratios of dominant frequencies have long been used to classify different orbital types and even orbital families, we use the frequencies $f_x$ and $f_y$ together and for all orbits in our models, without exception. Thus, we obtain an extended classification of orbits in the bar area. Next, we will examine how quantitative changes in populations of different orbital groups are associated with morphological changes, especially with barlens morphology. 

\subsection{Main orbital groups}
\label{sec:main_groups_bar}

\begin{figure*}
\centering
\includegraphics{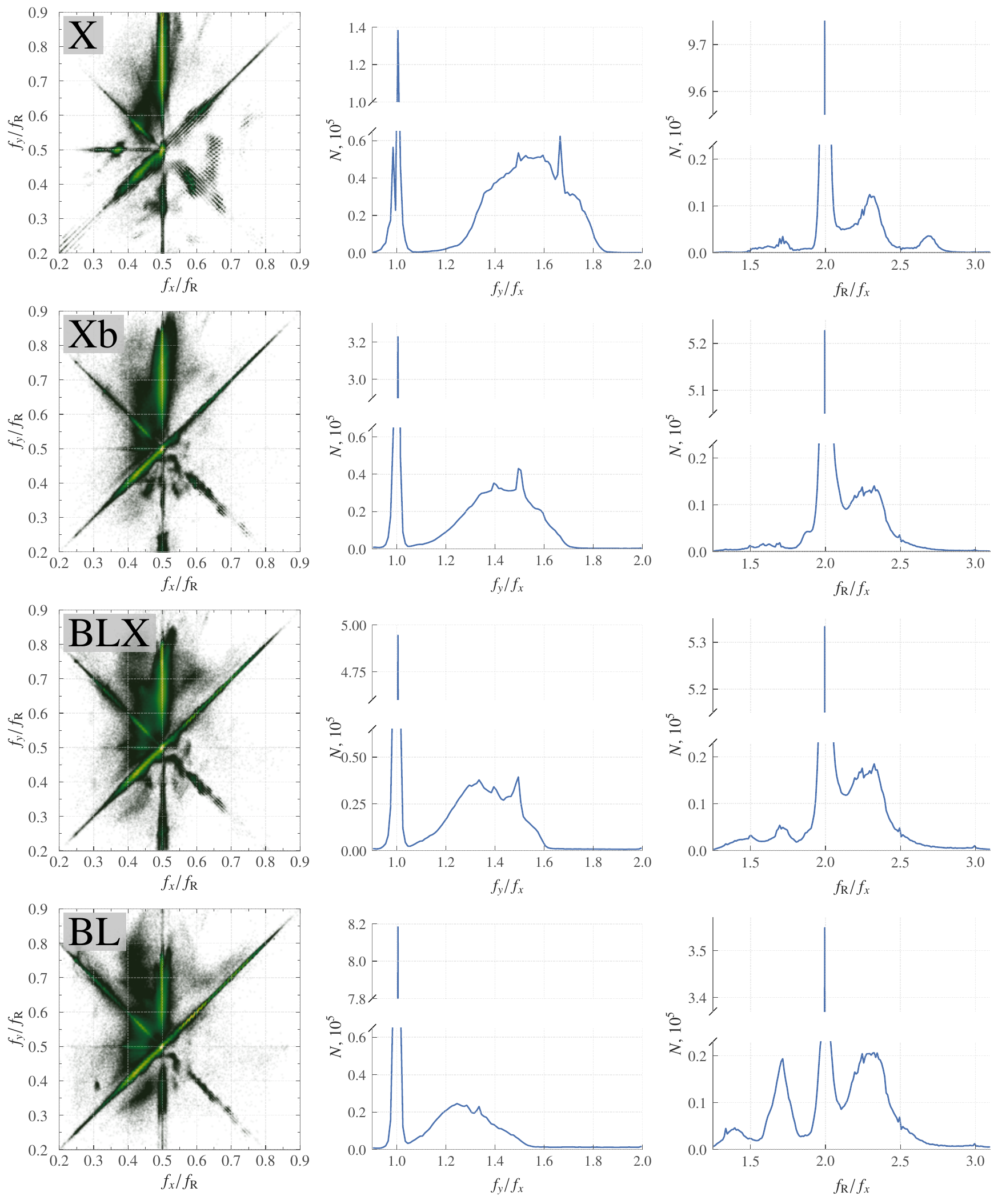}
\caption{Distribution of all orbits constituting the bar over the ratios of dominant frequencies for X, Xb, BLX and BL models (top down) calculated over the time interval $t=400-500$~t.~u. (5.3 -- 6.6 Gyr). {\it Left}: 2D maps on the plane $f_\mathrm{R}/f_x$--$f_y/f_x$. {\it Middle and right:} 1D distribution over the ratios $f_y/f_x$ and $f_\mathrm{R}/f_x$. Bin width is equal to $0.01$.}
\label{fig:fxyR_bar}
\end{figure*}

We start with a qualitative description of the main orbital blocks. Fig.~\ref{fig:fxyR_bar} (left column) shows 2D maps of the ratios of dominant frequencies, $f_x/f_\mathrm{R}$ and $f_y/f_\mathrm{R}$. In such coordinates, the maps demonstrate a curiously regular pattern across different models. One can see a bright spot at $(f_x/f_\mathrm{R};f_y/f_\mathrm{R}) = (0.5;0.5)$ and at least four straight rays going from it. The spot and the rays can be associated with the corresponding orbital groups. All these features are present on 2D maps for all models, regardless of their morphology.
\par
\begin{figure*}
\includegraphics{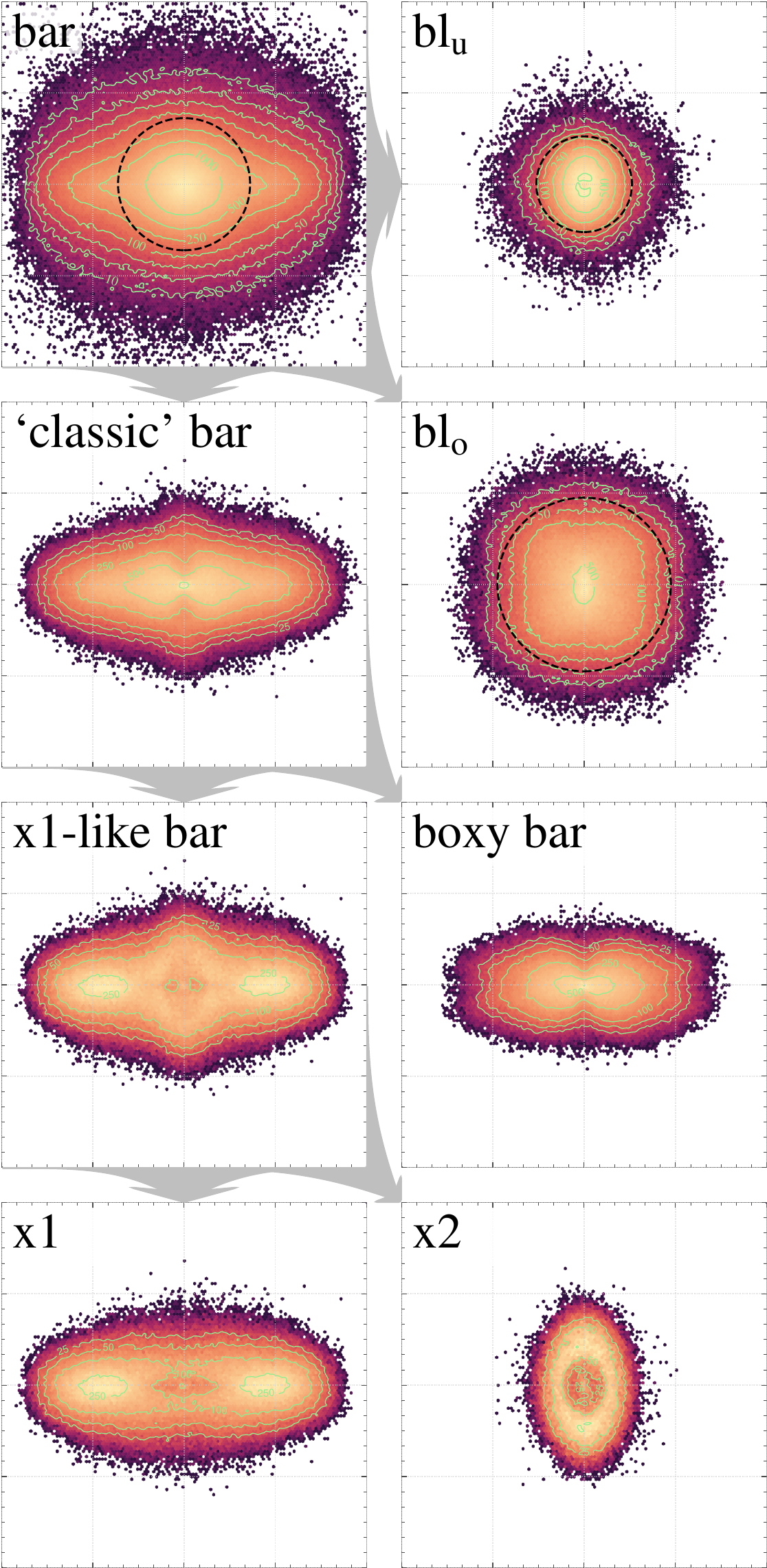}
\caption{Face-on view of all bar building blocks for model BL at $t=450$ t.~u. (6~Gyr). All subplots depict the same space area $(xy)=(-2,2)\times(-2,2)$. \textcolor{black}{Dashed black circles show the areas within two effective radii for the lens orbital groups: in the top left panel, a circle of $2R_e$ for a sum of \blu and \blo is depicted, while \blu and \blo panels show circles of $2R_e$ for the corresponding orbital groups.}}
\label{fig:xy_collage_BL}
\end{figure*}

For the spot and the vertical ray going upwards we have $f_x=0.5 f_\mathrm{R}$. This equality was a criterion for the bar identification in \citet{Portail_etal2015b} and \citet{Gajda_etal2016}. 
In our reference frame $2 \pi f_x \approx \Omega-\Omega_\mathrm{p}$, so the equality $f_X=0.5f_\mathrm{R}$ is practically equivalent to the ``classic'' condition for a bar $\Omega - \Omega_\mathrm{p} = \varkappa/2$. \textcolor{black}{Families x1 and x2, usually considered in the literature, satisfy this condition since they are usually elliptical orbits in which the star makes two radial oscillations in one revolution around the centre of the system.}
This means that orbits falling into the spot and the vertical ray are mainly periodic orbits inside the bar plus quasi-periodic orbits around them. These orbits precess \textcolor{black}{almost} synchronously with a bar and can constitute a bar in its ``classic'' meaning\footnote{A more detailed discussion of the types of orbits is given in Section~\ref{sec:orbits}.} \citep{Portail_etal2015b, Gajda_etal2016}.
They are located at the inner Lindblad resonance (ILR) \citep{Athanassoula2003}.
The face-on projection of this orbital group is presented in Section~\ref{sec:bar_anatomy} (Fig.~\ref{fig:xy_collage_BL}, ``classic'' bar). The plot is very similar to the ILR image from the work of \citet{Ceverino_Klypin2007} (their figure~10), identified by the usual condition 
$\Omega-\Omega_\mathrm{p} = \varkappa/2$. 
The use of the criterion to identify a bar based on Cartesian frequencies $f_\mathrm{R} \approx 2f_x$ instead of the usual condition $\Omega-\Omega_\mathrm{p} = \varkappa/2$ leads to the same bar morphology as if we used the latter criterion. Our bars look like narrow and elongated structures like those typically distinguished by a ``classic'' condition in studies of this kind.
\par 
Thus, the bar orbits lie along a prominent strip, going vertically upward from a bright spot at the centre of a plot (Fig.~\ref{fig:fxyR_bar}, {\it left}). For the orbits falling into the spot we have $f_y/f_x=1.0\pm0.1$ and $f_\mathrm{R}/f_x=2.0\pm0.1$. We refer to this group as an x1-like family\footnote{These are not pure x1 orbits, because they have a length in the vertical direction. Moreover, this family may contain orbits from the x2 family and other families of higher orders. For brevity, we call all of them ``x1-like orbits''.}.
The vertical strip is associated with another group with $f_y/f_x>1.1$ and $f_\mathrm{R}/f_x=2.0\pm0.1$. It is constituted by the so-called box-shaped and boxlet orbits (\citealp{Valluri_etal2016,Abbott_etal2017,Chaves-Velasquez_etal2017,Gajda_etal2016}; see also Section~\ref{sec:bar_orbits}). 
\par
Three more rays are noticeable on 2D frequency maps. Orbits falling into these rays do not contribute to the bar in its ``classic'' meaning because they do not precess synchronously with the bar\textcolor{black}{, that is, they do not have $f_x/f_\mathrm{R}\approx0.5$} and have either $f_x/f_\mathrm{R}\lessapprox0.5$, or $f_x/f_\mathrm{R}\gtrapprox0.5$. 
At the same time, particles from these rays are definitely not from the outer disc, because here we consider particles only inside the ring, which separates the bar from the outer disc.
The upper left ray begins near a central spot and have $f_\mathrm{R}/f_x>2.1$; $(f_x+f_y)/f_\mathrm{R}=1.0\pm0.1$. A bottom left ray have $f_\mathrm{R}/f_x>2.1$ and $f_y/f_x=1.0\pm0.1$. 
There is a third ray going up to the right and having $f_\mathrm{R}/f_x<1.9$ with $f_y/f_x=1.0\pm0.1$.
\par
Orbits with $f_\mathrm{R}/f_x>2.1$ were mentioned by \citet{Harsoula_Kalapotharakos2009}. They were also found in the $N$-body simulations by \citet{Gajda_etal2016}. In \citet{Gajda_etal2016}, these orbits were found to satisfy an additional condition  $(f_x+f_y)/f_\mathrm{R}=1$, that is, these are the orbits that fall into the upper left ray.
There is another ray with $f_\mathrm{R}/f_x>2.1$ and an additional condition $f_y/f_x=1.0$ (Fig~\ref{fig:fxyR_bar}, {\it left}, line running from bottom to top, from left to right up to the point $f_x/f_\mathrm{R}=0.5$; $f_y/f_\mathrm{R}=0.5$). This orbital group was not explicitly mentioned in the literature. 
\par
\citet{Gajda_etal2016} noted that orbits similar to those of the upper left ray are not elongated along the bar and does not support it. As will be shown below, both left rays contribute to the barlens. 
\par
Orbits from the third ray going up to the right were not explicitly described previously in the literature to our knowledge. They have $f_\mathrm{R}/f_x<1.9$ and $f_y/f_x=1.0$. A possible complementary branch along the line $(f_x+f_y)/f_\mathrm{R}=1.0$ is practically not populated in all our models. Thus, for $f_\mathrm{R}/f_x<1.9$ there is only one orbital group.
\par
We studied the shape of the face-on isophotes for all particles of both branches with $f_\mathrm{R}/f_x>2.1$ and did not find a significant difference between the upper and lower rays, except that the orbits of the upper ray are assembled into a more compact structure than the orbits of the lower ray. For this reason, we view both branches as sub-parts of the one building block of the entire bar morphology, that is, we will consider them only in conjunction with one another in further discussion. Here we should note that both of these branches are constituted by orbits with rather complex spectra. In some cases, spectra show two waves of comparable amplitude and in some cases, they do not. According to that, we believe that we need to perform a more rigorous analysis of orbits spectra to rightfully distinguish between these two orbital groups. In the future, we intend to enter into more subtle details related to the difference in the spectra of these two groups. But in this work we are mainly interested in the morphology of structures supported by orbits with different frequency ratios, therefore, our perhaps not so rigorous treatment of these orbits is more than justified.
\par
To simplify the further discussion, we introduce the following notation for these three groups of orbits: ``\blu'' for orbits from the upper right ray and ``\blo'' for orbits from \textbf{both} bottom left and upper left rays. As we will see below, the orbits from these groups are the most important contributors to a barlens structure and hence the notation ``bl'' follows. Subscripts ``$\mathrm{u}$'' and ``$\mathrm{o}$'' reflect the position of these groups on 2D frequency map (Fig.~\ref{fig:fRx_all}) with respect to x1-like and boxy orbits: ``$\mathrm{o}$'' stands for `above' (over) the bar ($f_\mathrm{R}:f_x > 2:1$) while ``$\mathrm{u}$'' stands for `beneath' (under) the bar ($f_\mathrm{R}:f_x < 2:1$). In practise we distinguish \blo and \blu groups by $f_\mathrm{R}:f_x > 2.1$ and $f_\mathrm{R}/f_x <1.9$ conditions.
\par
There are several other lines, but the number of particles along them is negligible compared the number of particles inside any previously mentioned ray or central spot and we exclude them from further discussion.

\subsection{Distributions over the ratio of the in-plane frequencies}

Fig.~\ref{fig:fxyR_bar} ({\it left} column) demonstrates what types of orbits contribute to the bar. At a qualitative level, all plots for all models look similar. Unlike left plots of Fig.~\ref{fig:fxyR_bar}, middle and right plots show 1D distributions over $f_y/f_x$ ({\it middle} column) and $f_\mathrm{R}/f_x$ ({\it right} column) ratios for each of our models and they give an idea of the quantitative population of the different orbital groups. It can be seen that \textcolor{black}{these distributions have more or less similar shapes for all models} with the exception of the BL model which is quite peculiar.
A comparative analysis of these distributions for different models will allow us to understand how the population of different orbital groups changes quantitatively from one model to another and to identify what type of orbits give rise to a barlens in BL model.
\par
First, we can see in Fig.~\ref{fig:fxyR_bar} ({\it middle}) that the second wide peak in the distribution over $f_y/f_x$ (between $1.0<f_y/f_x<2.0$) and the distribution as a whole shifts towards lower values of $f_y/f_x$ with an increase in CMC (from top to bottom). All four distributions show the presence of orbits, which can be quasi-periodic and have a frequency ratio $f_y/f_x$ expressed by a rational fraction (small peaks in the distributions in the region $1.0<f_y/f_x<2.0$). This ratio also becomes smaller when moving from the model X to the model BL: 5:3 for X model, 3:2 for Xb and BLX models and 4:3 for BL model.
Besides these changes, the distribution over $f_y/f_x$ becomes significantly less populated. We note that this part of the distribution is mainly consists of box-shaped orbits that form the vertical strips in Fig.~\ref{fig:fxyR_bar}, {\it left} (in $f_y/f_\mathrm{R}>0.5$ area, boxy bar). In Section~\ref{sec:comparison} we will see that the length of a boxy bar formed by orbits with lower values of $f_y/f_x$ is less than the length of a boxy bar consisting of orbits with larger frequency ratios. That is, the boxy bar becomes shorter and less bright with an increase in CMC, and this creates more favourable conditions for the manifestation of the central barlens.
\par
Secondly, the height of the peak at $f_\mathrm{R}/f_x=2.0$  (Fig.~\ref{fig:fxyR_bar}, {\it right}), which is populated by the orbits of the ``classic'' bar, decreases from the model X to the model BL. Moreover, the model without a classical bulge (model X) demonstrates one dominant family of orbits with $f_\mathrm{R}/f_x=2.0\pm0.1$ and a very high peak at this value. At the same time, in the model BL, these orbits account only for 22\% of all disc orbits versus 50\% of the orbits involved in the whole bar. That is, an elongated bar becomes less populated.
\par
Thirdly, orbits of novel types appear from model X to model BL, and they become quite numerous. They lie to the right and left of the peak at $f_\mathrm {R}/f_x$=2.0 (Fig.~\ref{fig:fxyR_bar}, {\it right}). We previously distinguished these two groups of orbits as \blu ($f_\mathrm {R}/f_x < 1.9$) and \blo ($f_\mathrm {R}/f_x > 2.1$) using 2D frequency ratios maps. 
The orbital group bl$_\mathrm{o}$ is present in all our models but it is not well populated in the model X. These orbits become more numerous in the BL model with barlens and they can contribute to the barlens.
\par
As to the \blu orbits, although models Xb and BLX have a bit more of such orbits than model X, there are still very few such orbits in these models. They manifest themselves quite prominently only in the model BL. We believe that these orbits make the main contribution to barlens morphology. 
\par
The analysis performed above showed there is probably more than one group of orbits that can be associated with a barlens (\blu and \blo orbits). That is, a barlens has its own orbital composition consisting of at least two different groups of orbits. But before studying the orbital composition of the barlens in detail, let us analyse in detail the structure of a ``classic'' bar usually extracted by the condition $f_\mathrm{R}/f_x=2.0\pm0.1$.

\section{The anatomy of a ``classic'' bar in BL model}
\label{sec:bar_anatomy}

\subsection{Two main families constituting a ``classic'' bar}

We identify a ``classic'' bar according to the condition  $f_\mathrm{R}/f_x=2.0\pm0.1$ as in \citet{Portail_etal2015b,Gajda_etal2016}. For the model with a barlens (BL model) this component is presented in Fig.~\ref{fig:xy_collage_BL} (``classic'' bar \textcolor{black}{panel}). This component looks like a very ordinary elongated bar, without any traces of the barlens. In its centre, an X-shaped (or rather peanut-shaped) morphology is observed. There are also two weak vertical protrusions at $x = 0$. 
\par
Two different orbital families can be distinguished in this component. In Fig.~\ref{fig:fxyR_bar}, {\it left} they are grouped near a bright spot at $f_x/f_\mathrm{R}=0.5$; $f_y/f_\mathrm{R}=0.5$ and along a vertical strip above this spot. These two group form two different orbital families: an x1-like bar and a boxy bar (see previous Section).
\par
Fig.~\ref{fig:xy_collage_BL} (\textcolor{black}{x1-like bar and boxy bar panels, respectively}) demonstrates snapshots for these two families at $t=450$, or ${\approx}$6~Gyr. We have checked that the morphology of both configurations does not change over a time interval from $t=400$ up to $t=500$ (5.3~--~6.6~Gyr). Both families are equally populated (approximately 10-11\% of all 4$kk$ particles are involved in each of these structures). But the structures formed by orbits from these families are strikingly different in morphology. An x1-like family gives a narrow and elongated bar with an inner bar-like structure superimposed on an outer bar and oriented perpendicular to it (Fig.~\ref{fig:xy_collage_BL}, x1-like bar). Apparently, the inner bar is formed by the orbits from the x2 family \citep{Contopoulos_Papayannopoulos1980,Athanassoula_etal1983}. Box-shaped orbits constitute a face-on peanut (Fig.~\ref{fig:xy_collage_BL}, boxy bar).

\subsection{An x1-like bar}
\begin{figure}
\includegraphics{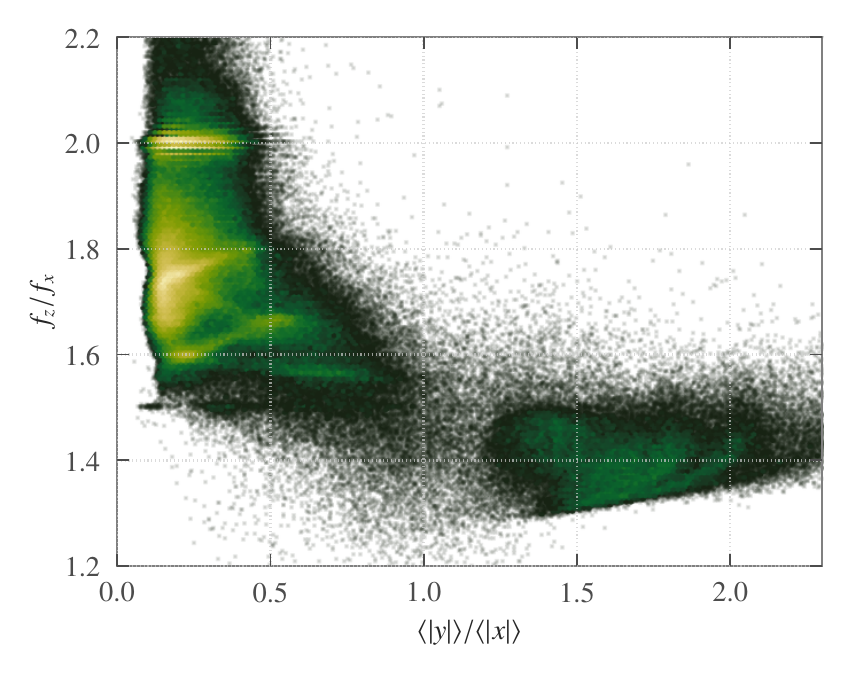}
\caption{2D distribution of x1-like particles over the ratios $f_z/f_x$ and $\averg{|y|}/\averg{|x|}$ for the model BL calculated over the time interval $t=400-500$.}
\label{fig:MyMx_2D_x1_BL}
\end{figure}

Two structures (a narrow and long outer bar and an inner perpendicular bar) in Fig.~\ref{fig:xy_collage_BL} (x1-like bar) can not be separated if we only use frequency ratios $f_\mathrm{R}/f_x$ and $f_y/f_x$. 
But if we look at 2D distribution of the ratios $f_z/f_x$ and $\averg{|y|}/\averg{|x|}$, where $\averg{|x|}$ and $\averg{|y|}$ are the mean absolute values of $x$ and $y$ coordinates of orbits calculated over time interval $t=400-500$, we will see two non-overlapping areas, where orbits contributing to the x1-like bar fall (Fig.~\ref{fig:MyMx_2D_x1_BL}). The boundary between two areas passes at $\averg{|y|}/\averg{|x|} = 1$. The ratio $\averg{|y|}/\averg{|x|}$ characterises the flattening of the orbit along the major axis of the bar. An orbit is elongated along the major axis of the bar if this ratio is less than one. It is almost round if the ratio is equal to one, and elongated along the minor axis of the bar if the ratio is greater than one. 
Thus, the area with $\averg{|y|}/\averg{|x|}$ is inhabited by the orbits elongated along the bar major axis. At the same time, these orbits have $f_z/f_x>1.5$. In the second area, orbits have $\averg{|y|}/\averg{|x|}>1.1$ and $f_z/f_x<1.5$. 
\citet{Portail_etal2015b} and \citet{Parul_etal2020} argued that the smaller $f_z/f_x$ are, the less elongated the orbits are. And this is what we see in Fig.~\ref{fig:MyMx_2D_x1_BL}. It can be assumed that the first orbital group contribute to a narrow and extended bar and the second one constitutes the innermost bump in Fig.~\ref{fig:xy_collage_BL} (x1-like bar). Moreover, the first group shows an enhancement at $f_z:f_x=2:1$. These are banana-like orbits that delineate the most remote parts of such a bar \citep{Parul_etal2020}. 
There is a long tail of orbits with $\averg{|y|}/\averg{|x|} \geq 1$, i.e. of orbits elongated along the bar minor axis. Apparently, these are orbits ``genetically'' related to the planar x2 family, but there are very few of them. 
\par
Snapshots for the individual orbital groups are presented in Fig.~\ref{fig:xy_collage_BL} (x1 and x2)
The right plot demonstrates a structure that is assembled from x2 orbits. 
Although this orbital family has been known for a long time \citep{Contopoulos_Papayannopoulos1980,Athanassoula_etal1983}, it seems that for the first time we clearly \textcolor{black}{distinguished this family as a whole structure by means of frequency analysis.}
In the left plot, one can see only an extended narrow bar. But neither the boxy-/peanut-shaped part of the bar, nor the x2 part of the bar, nor the elongated `stick' give even a hint of the barlens presence. Orbits supporting a roundish barlens must be sought among other orbital groups that we differentiate by the frequency ratios.

\section{The anatomy of a barlens}
\label{sec:barlens_anatomy}

\subsection{Main orbital blocks contributing to the barlens}

We have identified three orbital groups that do not enter into x1, x2 or box-shaped families constituting the bar. They form three rays in Fig.~\ref{fig:fxyR_bar} ({\it left}) coming from the bright spot at the point (0.5,0.5) left-down, left-up and right-up. We previously introduced them as bl$_\mathrm{u}$ and bl$_\mathrm{o}$ orbit groups (see Section~\ref{sec:plane_res}). 
The orbits that fall into these groups can contribute to the barlens. To support this hypothesis, we analyse the distribution of the average flattening of the orbits.
\par
\begin{figure}
\hspace*{-1ex}
\includegraphics{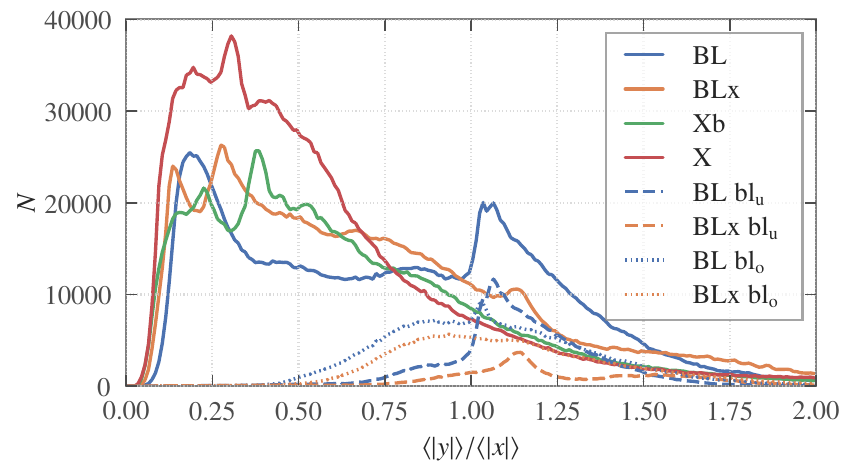}
\caption{1D distribution of all bar particles over the ratio $\averg{|y|}/\averg{|x|}$ for X, Xb, BLX and BL models calculated for the time interval $t=400-500$ (5.3 -- 6.6 Gyr). Bin width is equal to $0.01$. The dashed curves show the distribution of the same ratio but only for the bl$_\mathrm{u}$ orbits with $f_\mathrm{R}/f_x<1.9$ while the dotted lines are for bl$_\mathrm{o}$ orbits with $f_\mathrm{R}/f_x>2.1$.}
\label{fig:MyMx_1D_bar}
\end{figure}

Fig.~\ref{fig:MyMx_1D_bar} shows the distribution over the ratio $\averg{|y|}/\averg{|x|}$. It can be seen that although there is a certain number of orbits with the ratio
$\averg{|y|}/\averg{|x|} >1$ in the model X, the number of these orbits is very small. There are more of them in models Xb and BLX, and this is due to the appearance of the \blo orbits with $f_\mathrm{R}/f_x>2.1$. And when we look at such a distribution for model BL, we observe the appearance of a large second peak in the distribution over $\averg{|y|}/\averg{|x|}$ near the ratio 1.05-1.10. 
There is only a hint of such a peak in BLX model (the model with traces of a barlens), but for model BL it becomes very noticeable\footnote{In the BL model the initially identified bl$_\mathrm{u}$ group contains $9.97\%$ of orbits from 4$kk$. We found that a small number of such orbits ($0.75\%$ from 4$kk$) are elongated and have $\averg{|y|}/\averg{|x|}<0.5$. In the Appendix~\ref{sec:surgery}, we describe an algorithm for getting rid of such orbits. As a result, we left in this group $9.22\%$ of orbits from 4$kk$.}.
And it is in this model that a new type of orbits with $f_\mathrm{R}/f_x<1.9$ appears (\blu orbits) and the peak is mainly constituted by these \blu orbits (see the dashed lines in Fig.~\ref{fig:MyMx_1D_bar}). The orbits of this type are expected to have a roundish shape and, apparently, make a decisive contribution to the barlens.
\par
In Fig.~\ref{fig:xy_collage_BL} we plot a  snapshot depicting only \blu orbits at $t=450$ (6~Gyr). We also note that the entire configuration does not change its shape over a period of time from $t=400$ to $t=500$. Rounded isophotes and a rather strong concentration of the matter towards the centre are striking. 
\par
There is another source from which orbits supporting the barlens morphology are drawn. These are the \blo orbits with $f_\mathrm{R}/f_x>2.1$ (Fig.~\ref{fig:fxyR_bar}, right plots, to the right of the peak at $f_\mathrm{R}/f_x=2.0$), or $f_x/f_\mathrm{R}<0.5$ (Fig.~\ref{fig:fxyR_bar}, to the left of the vertical strip at $f_x/f_\mathrm{R}=0.5$). As we have mentioned, this group contains two branches: one with $(f_x+f_y)/f_\mathrm{R}=1$ and the other one with $f_y/f_x=1$ (Fig.~\ref{fig:fxyR_bar}, {\it left}). The structures formed by the orbits of these two branches differ little morphologically. One difference is that the latter branch contains a larger number of elongated and flattened orbits. The distribution of all \blo orbits over the ratio $\averg{|y|}/\averg{|x|}$ for BLX and BL models shows that a small number of orbits are very elongated along the major axis of the bar. In total, 17.42\% of 4$kk$ particles fall into the \blo group. After removal\footnote{See Appendix~\ref{sec:surgery}} of very elongated orbits,  
13.65\% remained. The distribution over the ratio $\averg{|y|}/\averg{|x|}$ for the remained orbits is shown in Fig.~\ref{fig:MyMx_1D_bar} (the dotted lines). The distribution has no prominent peak near $\averg{|y|}/\averg{|x|} \approx 1.1$. It rather wide but the structure associated with the \blo orbits shows barlens-like morphology (Fig.~\ref{fig:xy_collage_BL}, \blo plot).
\par
Without a doubt, the \blo orbits definitely contribute to the barlens, but the morphology of the structure formed by these orbits is rather unusual. In contrast to the structure shown in Fig.~\ref{fig:xy_collage_BL} ($bl_\mathrm{u}$) that has rounded isophotes, the isophotes of the structure formed by the \blo orbits has a square-like shape (see Fig.~\ref{fig:xy_collage_BL}, \blo subplot). We will discuss orbits capable of supporting such an unusual morphology in Section~\ref{sec:orbits}.
\par
Thus, it turns out that the barlens is a complex structure. The contributions of the rounded and square-shaped structures to the central barlens are comparable in the BL model ($9.25\%$ and $13.65\%$). \textcolor{black}{But, as indicated by eye, they have quite different density distributions. To make it more transparent, we calculated effective radii $R_e$ of the structures formed by \blu and \blo orbits, respectively, where $R_e$ is defined as the radius that contains half of the total number of particles in a given subsystem.
Quantity 2$R_e$ then can be considered a measure of the size of the structure. We found that the sizes of both structures are indeed quite different. For \blu, it is about 0.5 length units, while for \blo it is about 1.0 length units, that is, two times larger. The whole barlens, which is the sum of \blu and \blo groups, has an intermediate size, about 0.75 length units.} Thus, the morphology of combined \blo and \blu parts depends on the distribution of matter in each structure, on how loose or concentrated it is.  We will discuss this issue in the next Section~\ref{sec:comparison}. Concerning observational data, we would like to note that although usually by a barlens we mean something that has circular isophotes, like those in NGC~1015, NGC~1533, NGC~4394, NGC~4596, NGC~4608, NGC~4643, NGC~5101 galaxies (\citealp{Laurikainen_Salo2017}, Fig.~B.1), sometimes a barlens can have a square-shaped morphology. \textcolor{black}{Examples of galaxies with such unusual square-shaped barlens morphology are NGC~1022, NGC~3892 and NGC~5339 (\citealp{Laurikainen_Salo2017}, Fig.~C.2, C.9 and C.17). These galaxies are visible at low inclinations, almost face-on, and the boxiness of isophotes, especially in the region adjacent to the barlens, is not due to the thick part of the bar (\citealp{Laurikainen_Salo2017}, Fig.~5).}
\par 
\par 

\section{Comparative analysis of all models}
\label{sec:comparison}

\begin{figure}
\hspace*{-1ex}
\includegraphics{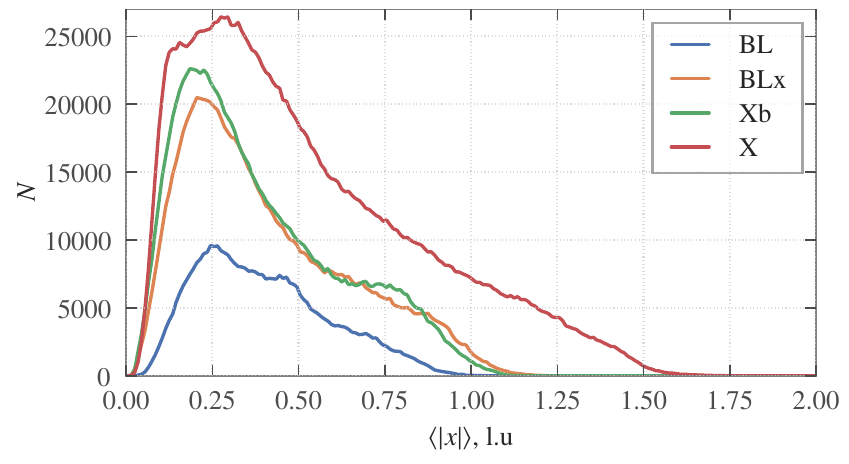}
\caption{1D distribution of the boxy bar particles over the value $\averg{|x|}$ for BL, BLX, Xb and X models calculated over the time interval $t=400-500$~t.~u. (5.3 -- 6.6 Gyr). Bin width is equal to $0.01$.}
\label{fig:Mx_1D_bar}
\end{figure}

Let us summarize the trends in the dominance of different types of orbits in our models in order to understand the reasons for the difference in morphology.
Table~\ref{tab:familiesnumbers} shows how, with the transition from a peanut-shaped bar (X) to a barlens (BL), the contribution of different types of orbits to the overall bar changes. Several trends are clearly visible. From the model X to the model BL, the number of boxy bar orbits decreases and the population of the x1-like bar family with $f_y/f_x{\approx}1$ grows on the contrary. If for a model without a bulge (model X) the bar mainly consists of box-shaped orbits (45\% from 4$kk$) with a very small admixture of x1-like orbits (4\%), then in the model with the CMC (model BL) the contribution of both orbital families is the same (10.5\% and 11\%, respectively). Thus, judging by our models, the orbits forming a narrow and long bar become significant only in potentials with a central spherically symmetric component. Next, while passing to the model BL, boxy orbits become less elongated along the major axis of the bar. This fact is reflected in the distribution over the mean absolute value of the $x$ coordinate $\averg{|x|}$ of boxy bar orbits ($f_y/f_x>1.0$, $f_\mathrm{R}/f_x=2.0\pm0.1$) calculated over the time interval $t=400-500$ (Fig.~\ref{fig:Mx_1D_bar}). When passing from the model X to the model BL, orbits with large values of this parameter leave the distribution and the size of the box-shaped bar is almost halved. 
\par
The proportion of orbits that do not support the bar (\blu and bl$_\mathrm{o}$ orbits) increases from the model X to the model BL. There is a significant difference in how these orbits manifest themselves in our models. The \blo orbits are present even in the X model (5\%). It is not surprising that such orbits were found in studies of this kind even in models without CMC (for example, \citealp{Gajda_etal2016}).
The contribution of the \blu orbits becomes significant only for the model BL (9\%). Their number is negligible even in the BLX model with a classical bulge of not too high concentration (3\%). This explains the fact that so far little attention has been paid to such orbits. They manifest themselves only in models with special conditions. And it is in these models that the central barlens appears, superimposed on the bar (model BL). 
\par
Despite the different populations of the \blu and \blo groups in different models, orbits that do not support the bar remain approximately constant in size. The decrease in the size of the structure formed by box-shaped orbits, and the invariance of the length of the structure supported by the \blu and \blo orbits, leads to the following consequences. As we move to the model with a barlens (BL), the family of the box-shaped bar ``sinks'' in the remaining orbits, and the peanut-like structure ceases to be visible, giving way to a rounded shape clearly distinguishable even against the background of a bar.

\begin{table}
\centering
\begin{tabular}{lr|rrrrr}
\hline
& {orbital}   & {X} & {Xb} & {BLX} & {BL} & \\
& {group}   &  &  &  &  & \\
\hline
\hline
& bar                  & 60.08 & 45.90 & 52.86 & 49.76 & \\
\hline
bar: & $\text{bl}_\text{o}$ & 5.36  & 7.82  & 9.98  & 13.49 & \\
& $\text{bl}_\text{u}$ & 0.96  & 1.05  & 3.42  & 9.22  & \\
& ``classic'' bar                  & 50.40 & 32.98 & 34.16 & 22.37 & \\
\hline
\hline
`classic bar': & boxy bar             & 45.19 & 25.68 & 23.88 & 10.45 & \\
& x1-like bar            & 3.79  & 6.52 & 9.51  & 11.35 & \\
\hline
\hline
x1-like bar: & x1                & 3.48  & 5.70 & 7.60  & 9.64  & \\
& x2                & 0.30  & 0.83 & 1.91  & 1.71  & \\
\hline
\end{tabular}
\caption{The percentage of orbits of each type in the models. The fraction is given relative to the total number of particles in the disc (4$kk$). Below the double line, there are the subgroups, into which the main group is divided just above the line. A slight mismatch in the amounts arises due to the ``cleaning'' procedure applied to the \blu and \blo orbits and a small number of irregular orbits with $f_y<f_x$ in the bar.}
\label{tab:familiesnumbers}
\end{table}
\par
\begin{figure*}
\centering
\includegraphics[width=0.9\textwidth]{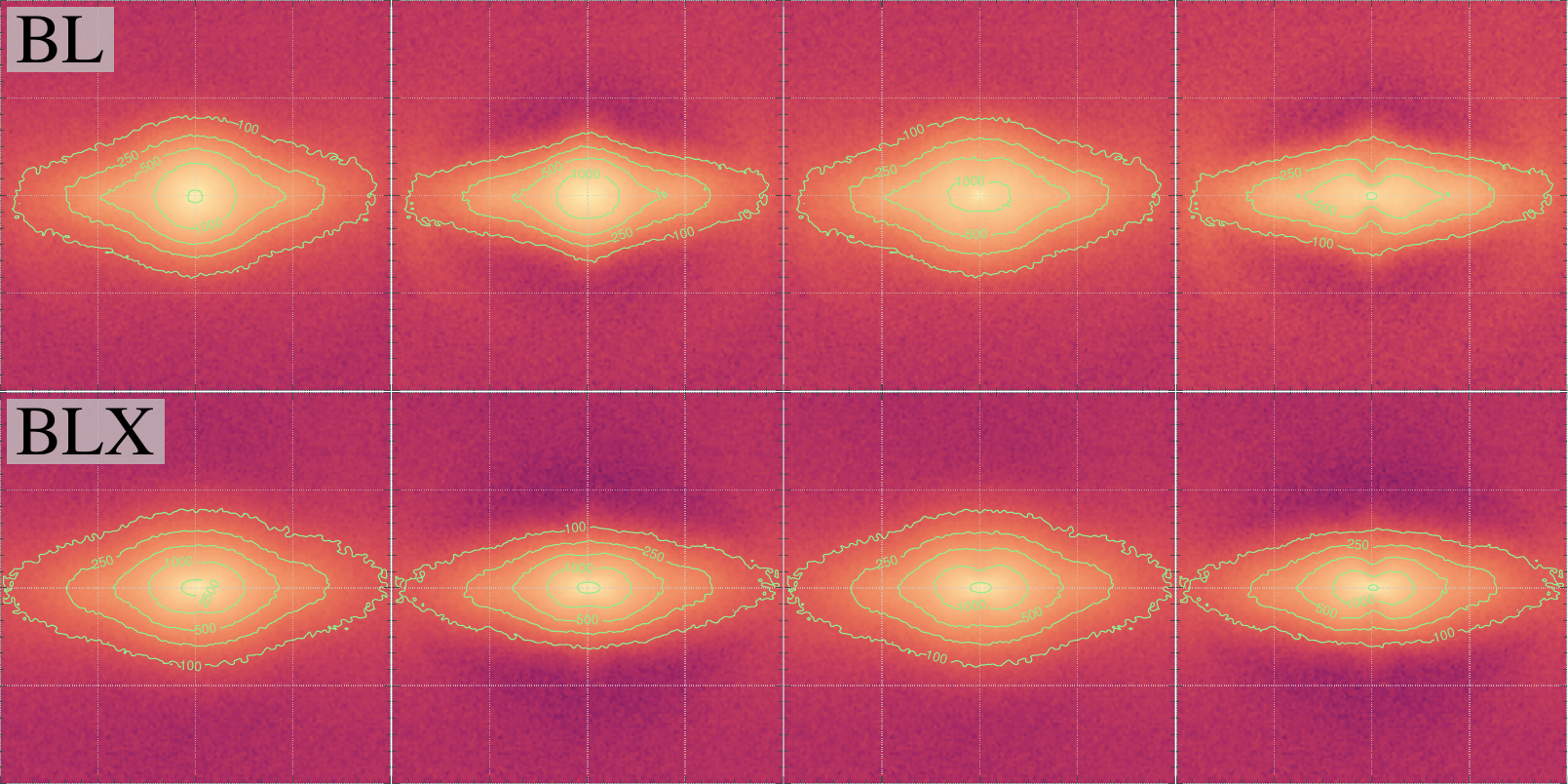}
\caption{Barlens ``disassembling'' into different orbital groups. {\it Top row} --- BL model, {\it botom row} --- BLX model. From {\it left} to {\it right}: all particles in the model, without \blo orbits; without \blu; without \blu and \blo. 
All plots depict the same square $(xy)=(-2,2)\times(-2,2)$ at $t=450$ (6~Gyr).}
\label{fig:barlensbuilding}
\end{figure*}

Both found orbital groups with $f_\mathrm{R} \neq 2f_x$ (bl$_\mathrm{u}$ and bl$_\mathrm{o}$) contribute to the morphology of a barlens, but their contribution is different. And this is due not only to the different populations of these groups. A good illustration of their different role in the barlens building is Fig.~\ref{fig:barlensbuilding}. In this figure, orbits that do not support the bar (bl$_\mathrm{u}$, bl$_\mathrm{o}$) are gradually excluded from BL and BLX models. In Fig.~\ref{fig:barlensbuilding} one can see how each of the orbital groups affects morphology. For model BL, after exclusion of the \blo orbits (13.5\%), the central barlens is still visible, although its overall length is getting smaller.
With the exception of only \blu orbits (9\%), the innermost isophotes lose their rounded shape. In the centre, there is even a hint of a peanut-like shape. However, outer isophotes still have a round shape, which is an indicator that there is a part of a barlens remaining. Only with the exception of both orbital groups, does the bar acquire a peanut-like shape. For BLX model, the exclusion of the \blu orbits (only 3.4\%) immediately leads to a peanut-like shape, although some rounded outermost isophotes are still preserved. The exclusion of \blo orbits results in the following effect. Even the outermost isophotes begin to bend towards the centre along the line perpendicular to the bar. Thus, the \blu orbits are responsible for the roundness of the inner isophotes, and the \blo orbits give elongation to the barlens in the direction perpendicular to the bar.

\section{Typical orbits constituting the main blocks of a combined bar}
\label{sec:orbits}

\subsection{Bar supporting orbits}
\label{sec:bar_orbits}

If we want to understand the physical reasons for the manifestation of different building blocks of a bar, the first thing to do is to examine what types of orbits constitute these blocks. The orbital analysis provides candidates for the backbone of the structure. We found such candidates by sorting all orbits based on frequency ratios $f_\mathrm{R}/f_x$ and $f_y/f_x$. For further analysis, it is very important to indicate the stable periodic or quasi-periodic orbits which constitute the backbone of the structures. These orbits can trap regular orbits around them. The whole configuration supported by such orbits will retain its shape over many revolutions.
\par
Numerous studies of analytical models of bars and bars in $N$-body models show that the main type of orbits supporting a bar in the disc plane is tube-shaped orbits rotating around the $z$-axis and elongated along the major axis of the bar \citep{Athanassoula2003}. According to the nomenclature introduced in \citep{Contopoulos_Papayannopoulos1980,Athanassoula_etal1983}, it is customary to refer to such orbits as x1 family. The orbits from this family are in 2:1 resonance and they have two radial oscillations per one revolution around the centre. 
Another family of orbits in 2:1 resonance is the so-called x2 family \citep{Contopoulos_Papayannopoulos1980}. Orbits of this family are elongated perpendicular to the bar. 
\par
In the 3D case, when we also have to take into account vertical perturbations, the situation becomes more complicated. New orbital families are generated by bifurcation from x1, x2 and planar families of higher order \citep{Pfenniger_1984,Skokos_etal2002a,Skokos_etal2002b,Harsoula_Kalapotharakos2009,Patsis_Katsanikas2014a,Patsis_Katsanikas2014b}. In this case, we need to consider the orbits of $x_1 v_i$ ($i = 1, 2, 3, ...$) type bifurcated from the initially flat orbit at the vertical resonances $f_z/f_x = 2, 3, ...$ \citep{Pfenniger_Friedli1991,Skokos_etal2002a}. 
The whole sets of $x_1 v_i$ and $x_2 v_i$ families are usually referred as x1 and x2 orbital trees, respectively. 
However, we are not interested in the vertical structure of the B/P bulge in this work, therefore, we do not differentiate planar and not planar orbits as well as periodic and quasi-periodic orbits trapped around stable orbits and refer to them as a whole using ``x1'' and ``x2'' notations, respectively.
\par
All these orbits are well known. They structure the phase space and create the backbone of the elongated bar and the inner perpendicular bar (Fig.~\ref{fig:xy_collage_BL}, x1 and x2 plots).
\par
The structure like a face-on peanut (Fig.~\ref{fig:xy_collage_BL}, a boxy bar plot) is built by orbits of another type. In addition to the ``tubes'', our models include a significant number of box-like orbits. \citet{Valluri_etal2016} examined a sample of 10,000 orbits in each of their two $N$-body models with bars and classified the orbits using a method based on frequency analysis. They considered only those orbits for which the radial oscillation frequency and the tangential frequency are in 2:1 resonance in a reference frame rotating with a bar pattern speed. \citet{Valluri_etal2016} connected the majority of orbits with the quasi-periodic box-shaped orbit in a rotating triaxial potential. Like 3D orbits elongated along a bar, box-shaped orbits are parented from the x1 family. Based on their $N$-body simulations, \citet{Valluri_etal2016,Abbott_etal2017,Chaves-Velasquez_etal2017} give a lot of examples of 3D non-periodic orbits in their $N$-body models, which have peanut-shaped projections both in their face-on and side-on views. Combined together these orbits can constitute the structure like that in Fig.~\ref{fig:xy_collage_BL} (a boxy bar plot). \citet{Valluri_etal2016,Abbott_etal2017,Gajda_etal2016} also found the so-called resonant boxlet orbits (`fish/pretzels', `brezels'). Such orbits can be a backbone of a face-on `peanut' structure. 

\subsection{Lens supporting orbits}

\subsubsection{The bl$_\mathrm{o}$ orbits}

Frequency analysis of $N$-body models revealed also non-bar orbital groups in central parts of discs \citep{MartinezValpuesta_etal2006,Voglis_etal2007,Wozniak_Michel-Dansac2009,Harsoula_Kalapotharakos2009}. A considerable number of particles having $(\Omega - \Omega_\mathrm{p})/\kappa < 0.5$ and $(\Omega - \Omega_\mathrm{p})/\kappa > 0.33$, that is, lying in between 2:1 and 3:1 planar resonances near $(\Omega - \Omega_\mathrm{p})/\kappa \simeq 0.4-0.44$, were found by \citet{MartinezValpuesta_etal2006} (figure~10, top left plot), \citet{Voglis_etal2007} (figure~21, group A), \citet{Wozniak_Michel-Dansac2009} (figure~8,9, top), \citet{Harsoula_Kalapotharakos2009} (figure~6, right plots, groups A and B). All these orbital groups fall into a group of orbits that we introduced as \blo in the present work. These orbits have a frequency ratio $f_\mathrm{R}/f_x$ in between 2.1 and at least 3.0. They produce square-like morphology when depicted all together (Fig.~\ref{fig:xy_collage_BL}). \citet{Gajda_etal2016} described an orbital group with a `square' morphology with $f_\mathrm{R}/f_x \approx 2.27$ and $(f_x+f_y)/f_\mathrm{R}=1$. It is only a part of the upper left ray in Fig.~\ref{fig:fxyR_bar}, {\it left}. Thus, having analysed the frequencies for all, and not just for some pre-selected orbits, we determined all the orbits of this type, and not of individual representatives of this group falling into a specific region of the mentioned ray.
\par
\citet{Voglis_etal2007,Harsoula_Kalapotharakos2009} provide examples of regular orbits belonging to this group, including 5:2 resonant orbits. These orbits indeed have a `square' morphology (figure~22, 5:2, group A, \citealp{Voglis_etal2007}; figure~7, groups A,B \citealp{Harsoula_Kalapotharakos2009}). The authors have used time series for each orbit, the length of which corresponds to a time interval of hundreds of radial periods. In our models, such  orbits slowly precess at a speed lower than the angular speed of the bar and lag behind it, since $f_\mathrm{R}/f_x>2$.  The orbits of these type make less than 20 revolutions, without having time to draw a square in our case. However, an ensemble of particles with arbitrary initial phases has no problem to do it, which we precisely observe in Fig.~\ref{fig:xy_collage_BL}.
\citet{Gajda_etal2016} gave an example of a typical quasi-periodic orbit with $(f_x+f_y)/f_\mathrm{R}=1$ in their figure~6, row (c) and concluded that this type of orbit does not seem to support the bar. Such an orbit also resembles the orbit in figure~A6 in \citet{Patsis_Athanassoula2019}. The second bottom left ray of \blo group of orbits, that with $f_y/f_x=1$ (Fig.~\ref{fig:fxyR_bar}, {\it left}), was not explicitly identified in other works, but the orbits from this branch produce a square-like structure which is quite similar to that of the first branch according to our analysis. They differ only in that the orbits with $f_y/f_x=1$ are assembled into a more extended structure, while those with $(f_y + f_x)/f_\mathrm{R}=1$ seem to responsible for a more compact configuration. 

\subsubsection{The bl$_\mathrm{u}$ orbits}
A noticeable group (group B) of orbits $\kappa/(\Omega - \Omega_\mathrm{p}) < 1.67$ is visible in figure~21 in \citet{Voglis_etal2007}. Figure~22 \citep{Voglis_etal2007} shows that such orbits are \textcolor{black}{spherical on the plane of rotation and resemble} rosettes with many loops. 
In general, such orbits are undeservedly deprived of attention in the studies of the \textcolor{black}{$N$-body} bar internal structure.
A possible lack of interest in such orbits is due to the fact that this orbital group becomes noticeable only when a compact bulge is added to the $N$-body model, as is clearly visible in Fig.~\ref{fig:fxyR_bar}, {\it left} and in Table~\ref{tab:familiesnumbers}. \textcolor{black}{For example, we do not see such orbits in the frequency map in \citet{Harsoula_Kalapotharakos2009} (their figures~5,~6) while orbits from the group B  \citep{Voglis_etal2007} with its resonance members 5:3 can be associated with our \blu orbits.} 
\textcolor{black}{But even if such orbits are present in the models, they are little studied and \citet{Voglis_etal2007} do not discuss these orbits as building blocks for the barlens.} Apparently, this is the first time we have identified {\it all} the orbits of this type in $N$-body simulations \textcolor{black}{and regard them as the main orbits supporting the barlens.}
\par
Unlike the orbits lying in the bar, \blu orbits have a pronounced second peak in the periodograms $x(t)$ and $y(t)$. Moreover, for most of the orbits, in addition to the equality $f_y/f_x=1$, the ratio $f_y^{(1)}/f_x^{(1)}$ is also equal to 1, where $f_x^{(1)}$ and $f_y^{(1)}$ are frequencies of the secondary peaks. We also found that orbits from these group satisfy the following equalities:  $(f_x+f_x^{(1)})/f_\mathrm{R}=(f_y+f_x^{(1)})/f_\mathrm{R}=1$, $\varphi_y-\varphi_x = \pm \pi/2$, $\varphi_y^{(1)}-\varphi_x^{(1)} = \mp \pi/2$ where $\varphi_{x,y}$, $\varphi_{x,y}^{(1)}$ are initial phases. In general we found that most of the orbits of \blu group are fairly accurately described by sum of two oscillators along $x$-axis and sum of two oscillators along $y$-axis (see Eq.~\eqref{eq:rosette}), and such an orbit is a rosette \textcolor{black}{(a loop orbit, \citealp{Binney_Tremaine2008})}:
\begin{equation}
\begin{array}{lc}
x(t) = A_x \cos (2\pi f_x t + \varphi_x) + 
A_x^{(1)} \cos(2\pi f_x^{(1)} t + \varphi_x^{(1)}), & A_x > A_x^{(1)}; \\
y(t) = A_y \cos (2\pi f_y t + \varphi_y) + 
A_y^{(1)} \cos(2\pi f_y^{(1)} t + \varphi_y^{(1)}), & A_y > A_y^{(1)},
\end{array}
\label{eq:rosette}
\end{equation}
where $A_x$, $A_y$, $A_x^{(1)}$ and $A_y^{(1)}$ are the corresponding amplitudes.
\par
\begin{figure}
\begin{minipage}{0.5\textwidth}
\centering
\includegraphics[width=\textwidth]{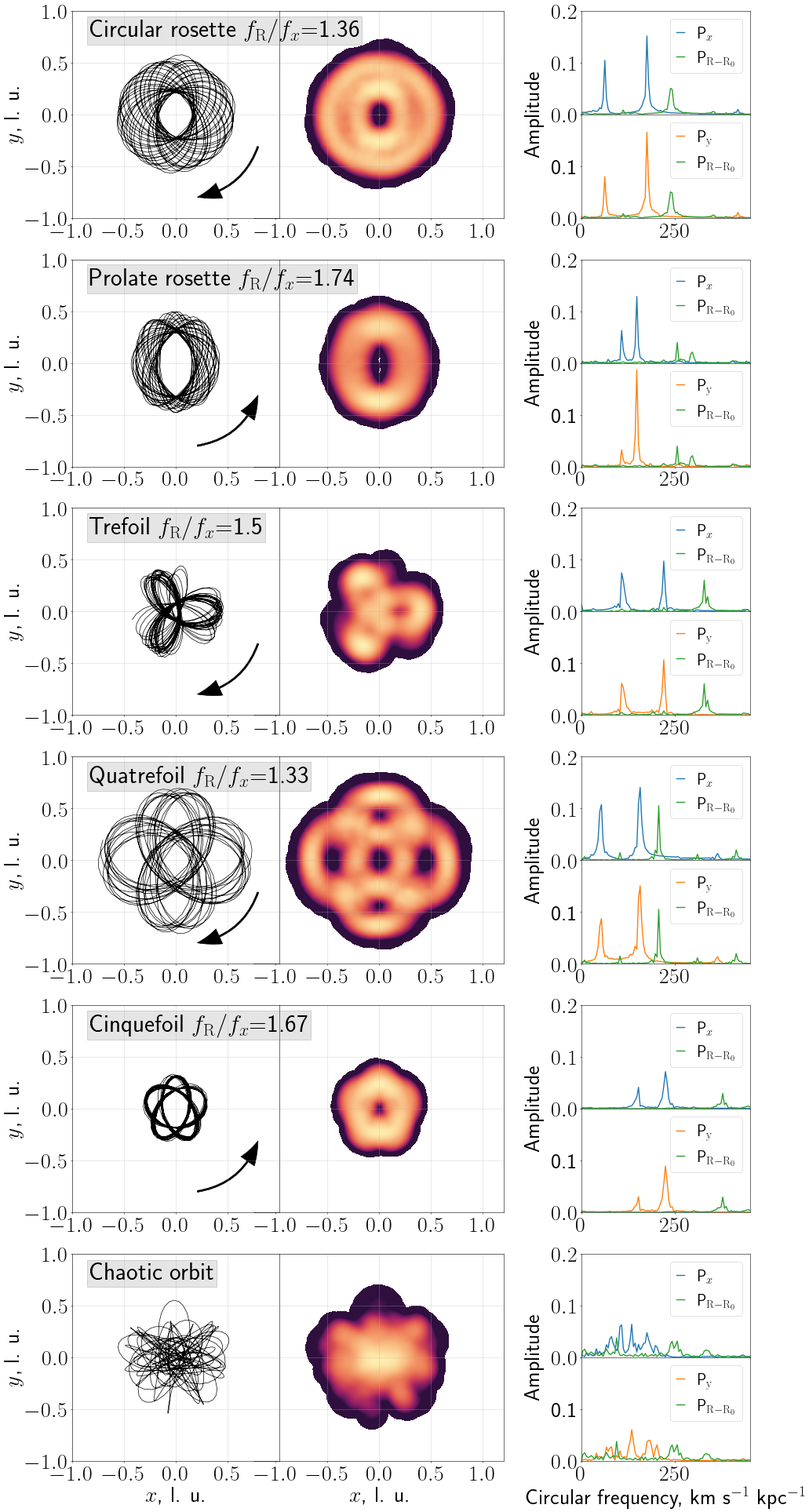}
\end{minipage}
\caption{Typical orbits constituting a barlens. \textit{Left:} $(xy)$ view of orbits, arrows indicate the direction of rotation (the bar rotates counter-clockwise); \textit{Middle}: the orbital density profile (see text for details); \textit{Right:} The corresponding spectra of oscillations along $x$ and $y$ axes plus a cylindrical radius $R$.}
\label{fig:orbits_all}
\end{figure}

For most of the orbits, the second frequency is less than the first one $f_{x,y}^{(1)} < f_{x,y}$. As a rule, such orbits precess quite quickly in the same direction in which the bar rotates. There are also very few orbits with $f_{x,y}^{(1)} > f_{x,y}$. These orbits usually precess very slowly in the opposite direction. Fig.~\ref{fig:orbits_all} shows different examples of bl$_\mathrm{u}$ orbits which we have found in the BL model. Typical rosettes (\textcolor{black}{loop orbits}) are shown in the upper two rows. \textcolor{black}{We note that the two displayed rosettes rotate in different directions implying that they originate from different orbital families. The roundish rosette resembles the quasi-periodic orbit around x4 orbital family while a prolate rosette seems to be a quasi-periodic orbit around x2 family.} The plots in the middle depict the density profile produced by each orbit over a long time interval ($\approx$ 1 Gyr). These plots are obtained in the following way. We colour each point with coordinates $(x, y)$ according to how many times a particle passes through this point during the considered time interval. Such plots directly reflect the morphology produced by a large number of the given orbits. The plots on the right are spectra of $x(t)$, $y(t)$ and $\mathrm{R}(t)$ oscillations.
\par
\par
The most of \textcolor{black}{\blu} orbits appear to be regular ones, \textcolor{black}{although we cannot reliably distinguish such orbits from sticky chaotic ones, since the time interval under consideration is relatively small.}
\textcolor{black}{Nevertheless, our statement about the regular nature of these orbits in the considered time interval is based on the fact that \blu group (as well as other groups) forms a pronounced line in the frequency ratios plane (Fig.~\ref{fig:fxyR_bar}) which implies that the regular ``backbone'' of orbits should be there (see~\citealt{Valluri_Merritt1998}).} Some of \blu orbits are slightly elongated along the minor axis of a bar. Among them, there are also $n$-foil type of orbits that have an appearance of \textcolor{black}{quasi-periodic} orbits~(rows 3-5 in  Fig.~\ref{fig:orbits_all}): trefoils with a frequency ratio 3:2, quatrefoils with 4:3 and cinquefoils with 5:4. They are not elongated either along the major axis or along the minor axis of the bar. It is natural that a mix of different types of such orbits plus the regular orbits around them that have rosette-like shape appears as a roundish structure, which we observe as a barlens. \textcolor{black}{We also note that the orbits presented are not planar but have a maximum vertical extension about 0.3 length units.}
\par
\textcolor{black}{\citet{Voglis_etal2007} give example of 3:1 and 4:1 orbits (their figure~22) which also have shapes of trefoils and quatrefoils. The 3:1 trefoils can be associated with a small peak in $f_\mathrm{R}/f_x$ plot (Fig.~\ref{fig:fxyR_bar}). Because $f_\mathrm{R}/f_x > 2.0$ for such orbits, they fall into the \blo group as well as 4:1 orbits. In our models these orbits are more extended than 3:2 trefoils from the \blu group. 3:1 and 4:1 orbits can also be described by the model of two oscillators but with $f_{x,y}^{(1)} > f_{x,y}$ ($A_{x,y} > A_{x,y}^{(1)}$).}
\par
Besides the regular orbits, \blu group contains some number of \textcolor{black}{clearly} chaotic orbits with a forest of lines in the spectra. An example of such an orbit is shown in Fig.~\ref{fig:orbits_all} (\textit{bottom row}). These orbits also can contribute to the roundish shape of a barlens. 
\par
We should also remind that all \blu orbits constitute only a part of a barlens. There is also \blo orbits that also seem to contribute to it.

\section{Discussion}
\label{sec:discussion}

\citet{Kormendy_2013} argues that lenses in many galaxies are defunct bars and suggests that bars evolve away into lenses due to the secular evolution that increases the central mass concentration so much that the bar orbits can no longer precess synchronously \citep{Kormendy_1979}. Some theoretical aspects of bar-to-lens evolution were reviewed in \citet{Combes_2008,Combes_2011a,Combes_2011b}. In these works, the appearance of the lens is associated with the gas inflow and gradual formation of the central mass concentrations (CMC). It is indeed well known that the addition of a CMCs like a central black hole leads to a bar weakening and even its possible destruction \citep{Norman_etal1996}. \citet{Combes_2011a} also emphasises an additional important aspect of bar destruction. Besides the gas inflow there is also an exchange of momentum between gas and bar. Gas gradually loses its momentum and gives it to the bar. Inside the co-rotation, the bar is a pattern with a negative angular momentum and this exchange makes the central orbits rounder and the bar weaker. Simulations show that the gas infall of 1-2\% of the disc mass is enough to transform a bar in a lens (e.g., \citealp{Berentzen_etal1998,Bournaud_Combes2002,Bournaud_etal2005}).
The formation of a central barlens on the background of a still existing bar \citep{Athanassoula_etal2015} is apparently associated with the same mechanism: partial destruction of the bar by growing CMC.
\par
\begin{figure*}
  \includegraphics{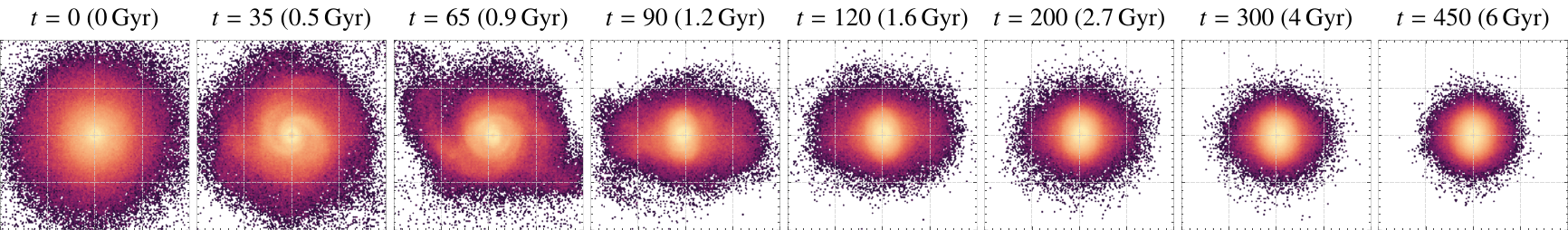}
  \caption{%
    Time evolution of a structure assembled from particles that make up \blu group captured at several moments
  from $0$ to 450 t.~u. (${\approx}$6 Gyr). Face-on views are displayed in the square $(xy) = (-2,2)\times(-2,2)$. 
}
  \label{fig:blu_timeevol}
\end{figure*}

\textcolor{black}{Our simulations describe a different situation, namely, not the destruction of an already formed bar by an increasing central density concentration, but the formation of a bar in the presence of an already existing density concentration. Since the axisymmetric part of the potential (a compact bulge) dominates in the central regions of the model (BL), and the forming bar introduces only a slight distortion, in the central regions we are dealing with the case of a weak bar and with non-closed loop orbits (rosettes) around closed loop orbits from the nearly round families x4 or x2 \citep{Binney_Tremaine2008}.}
We specifically traced the evolution of the barlens-like structure supported by \blu orbits from the very beginning of the simulations. We rewound it in time to the very beginning of the model evolution and it turned out that the barlens-like structure is formed in situ, and the influx of matter into this structure from other areas practically does not occur (see Fig.~\ref{fig:blu_timeevol}).
Thus, we can state that bl$_\mathrm{u}$ group of orbits \textcolor{black}{is formed naturally in the presence of  axisymmetric compact bulge without the need to destruct the bar component (or some part of it). 
}
\par
\textcolor{black}{We should also note that} barlenses are \textcolor{black}{morphologically} distinct from standard lenses (see the review by \citealp{Athanassoula2016}). The lens is a shallow feature on the brightness profile with further steeper declination \citep{Kormendy_1979}, while a barlens has a steep quasi-exponential profile \citep{Laurikainen_etal2014}.
\par
Disassembly of the barlens in our models into separate orbital groups with its own special morphology allows us to determine explicitly the surface density profile of a barlens and to offer an explanation for some observational features that are found in galaxies with barlenses. 
\par
\citet{Laurikainen_Salo2017} give examples of six galaxies with `bl' (barlens) in  their classification and a weak X-shaped feature in the unsharp-masked images. The most impressive examples are IC~1067 and NGC~4902 galaxies (\citealp{Laurikainen_Salo2017}, their figure~B.4, p.~54). The face-on views of models with the gas from \citet{Athanassoula_etal2015,Laurikainen_etal2014} demonstrate a barlens but almost all their models have traces of an X-like morphology. The same can be noted about the model by \citet{Salo_Laurikainen2017} with a low-mass bulge of the classical type ($B/T=0.01$) and the model by \citet{Smirnov_Sotnikova2018} with a massive bulge ($B/T=0.2$, figure~9). But \citet{Shen_Sellwood2004} noted that for a given central mass, dense objects cause the greatest destructive effect on the bar, while significantly more diffuse objects have a lesser effect. That is why the model by \citet{Salo_Laurikainen2017} with $B/T= 0.08$, but with the same effective radius as the model with $B/T=0.01$, leads to the formation of a strong barlens without traces of the face-on peanut. The same is true for our BLX and BL models.
\par
Since the manifestation of a particular morphology is determined by the delicate balance between the populations of different type orbits, we can assume that in galaxies with barlenses and traces of X-shaped structures, viewed  face-on, the central mass is not too compact and phase mixing has not yet occurred. The box-shaped part of the bar is squeezed with an increase in central density concentration, drowns in the orbital group forming the barlens (first of all, bl$_\mathrm{u}$), but still continues to shine through it.
\par
Another tiny feature of the barlens structure that almost no one paid attention to in the literature is as follows. The external isophotes of some barlenses show rather square outlines.
\textcolor{black}{One example is NGC~5339} galaxy from the sample by \citet{Laurikainen_Salo2017}. The authors depict \textcolor{black}{square-like isophotes with slightly rounded corners for this galaxy (figure~C.17, p.~89). The boxiness of isophotes was measured using the parameter $B_4$, which is associated with the $\cos 4\theta$ terms of the Fourier expansion of the isophotal shape. At a small inclination, NGC~5339 has strongly negative values of this parameter in the barlens region and falls out of the general trend in figure~5 \citep{Laurikainen_Salo2017}. Thus, the box-like isophotes for this galaxy can be associated with the face-on morphology (barlens) and not with the structure of the B/P bulge, which reveals itself at large inclinations.} 
We can only assume that the bl$_\mathrm{o}$ orbits with `square' morphology come to the fore in such galaxies, and the bl$_\mathrm{u}$ orbits are lost against its background or are confined in central regions.
\par
Table~\ref{tab:familiesnumbers} shows that in our models there are orbits, which we designated as x2. The population of this group of orbits increases with the transition to models with a central concentration. Apparently, this is a mixture of tube-shaped orbits of the x2-tree and x4-tree.
\textcolor{black}{One of the main differences} between x2 and x4 orbits is the 
sign of the angular momentum around the $z$ axis: the former rotate in the same direction as the bar in the inertial reference frame, and the latter rotate retrogradely \citep{Contopoulos_Papayannopoulos1980}. It is interesting to note that in the model X we found no x2 orbit, but there is a small number of x4 retrograde orbits. The same result was obtained in \citet{Valluri_etal2016,Voglis_etal2007}, who used models with initial conditions that did not include a classical bulge. Moreover, in all considered models with a bulge, the inverse relation is observed, and x2 orbits are much more populated than x4 orbits. \textcolor{black}{We want to emphasize that $N$-body simulations including gas usually form a nuclear ring (e.g. \citealp{Rautiainen_etal2002}). Such a ring is an indicator of the presence of the x2 family (see, for example, \citealp{Athanassoula_1992a,Athanassoula_1992b}). However, in pure stellar $N$-body simulations, such orbits are barely found. It is believed that structures perpendicular to the bar have been seen at the beginning of the model evolution  and then are quickly lost (e.g., \citealp{Shaw_etal1993}). In our case,  such a structure is a long-lived one.} In the BL model, the structure supported by these orbits has a large extent in the vertical direction. There is a great temptation to identify the structure supported by these orbits with an inner bar perpendicular to the main bar because galaxies with bulges tend to have double bars \citet{deLorenzo_etal2019}. However, even in the BL model, this structure has a large extent in the vertical direction and, apparently, makes an additional contribution to the barlens.
\par
\begin{figure}
  \hspace*{-1ex}\includegraphics{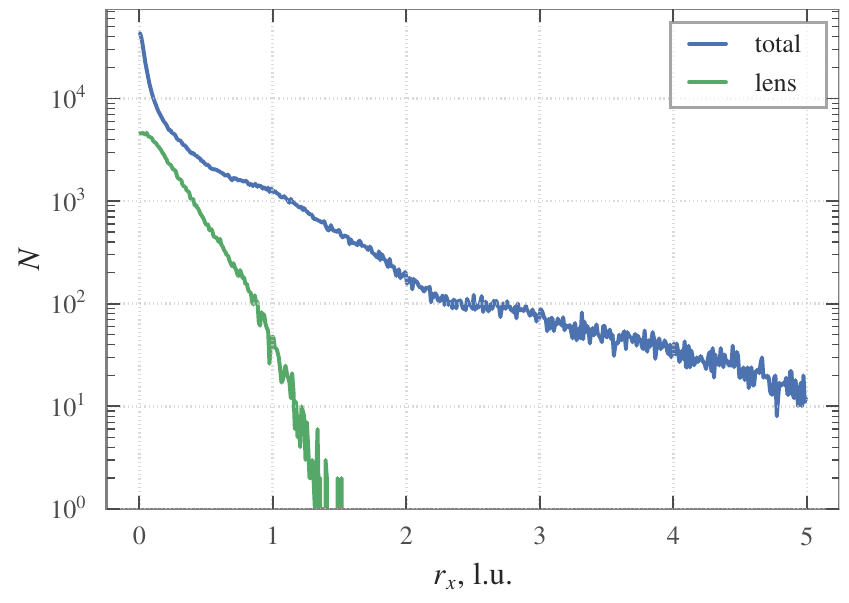}
  \caption{%
    Surface density radial profiles of the total model and barlens along the bar major axis. 
    The width of a slit is 0.12 length units.
  }%
  \label{fig:profiles}
\end{figure}

Identification of the barlens as a separate orbital structure helps to understand how it can contribute to the overall photometric profile. Unlike conventional photometric decomposition, for our models we can build a radial profile of the barlens as it is, without the need to account for other components. Fig.~\ref{fig:profiles} shows the surface density profile along the major axis of the bar for the BL model. The plot depicts an overall surface density profile, including a bulge, and an individual profile of the barlens, which is the sum of \blu and \blo orbit groups. An outer exponential disc is clearly visible on the overall profile. A little hump near $r_x=3$ coincides with the ring location, while the elongated bar gives a characteristic shoulder at $r_x{\approx}1$. As for the barlens, its surface brightness profile is very close to exponential. This is in good agreement with the results of the B/D/barlens/bar photometric decomposition of a number of galaxies \citep{Laurikainen_etal2018}, as well as numerical models with barlenses \citep{Salo_Laurikainen2017}: the surface brightness profiles of barlenses are nearly exponential. \textcolor{black}{Apparently, the structures built by \blu and \blo orbits are not related to inner, steep exponential discs, or discy bulges. Such inner exponential discs are described by \citet{Athanassoula2013}, referring to simulations with gas which concentrates to the inner parts of the disc and forms an inner disc. However, the extent of this region is of only of the order of 1~kpc
or even smaller. At the same time, a forming barlens is much more extended structure. \citet{Athanassoula_etal2015} also consider it unlikely that a discy bulge could be mistaken for a barlens. There are many reasons that these are different subsystems, the most important of which are differences in size, and that stellar populations of discy bulges are on average younger than that of barlenses.}

\section{Conclusions}
\label{sec:conclusions}

We analysed the orbital composition of the stellar disc for four $N$-body models with bars that differ in mass and scales of the initial classical bulge. We focused on the morphology of the orbits in the disc plane. In full agreement with the results described in \citet{Salo_Laurikainen2017}, we found that as the central concentration of the bulge increases, the morphology of the face-on bar changes. A smooth transition occurs from the peanut-shaped bar to the bar with a barlens.
\par
We translated the results of our $N$-body simulations into the ``language'' of orbits and suggested a solution to a number of dynamic phenomena. 
To investigate the orbital structure of the studied models, we used the frequency analysis method pioneered by \citet{Binney_Spergel1982}. 
We applied FFT analysis to all particle orbits ($N$ = 4$kk$) of the stellar discs of our models. 
\textcolor{black}{We proceed from the assumption that it is the quasi-periodic orbits that determine the long-lived structures in the stellar disc. The spectra of such orbits should consist of a discrete set of peaks.}
We calculated two main frequencies from the time series of Cartesian coordinates and a cylindrical radius for all of the orbits in the disc. 
Work of such a volume of studying the morphology of orbits in a bar before this was carried out only in \citet{Parul_etal2020}. 
Based on the obtained frequencies and their ratios, we have identified all particles that contribute to the entire bar in our models. 
Further, a classification scheme for different types of orbits based on two in-plane frequency ratios ($f_x/f_\mathrm{R}$ and $f_y/f_\mathrm{R}$) was extended and applied to all particles in the area of the bar. The analysis of ratios of different frequencies allowed us to distinguish all orbits that structure the phase space and are responsible for the morphological features of the so-called B/P bulges in our models.
These orbits include those that are not elongated along the bar and constitute the barlens.
\par
A comparison of the populations of different types of orbits in different models showed that the peanut-shaped morphology of the bar is created mainly by box-shaped orbits located at the inner Lindblad resonance. This confirms the results of previous studies, based both on the analysis of a large number of orbits in ``frozen'' potentials and on the analysis of a sample of orbits in $N$-body potentials.
\par
Using the ``orbital language'', we have proposed an explanation of the difference between the two types of bars that stand out in a face-on view, barlenses and peanut-shaped bars. Although our models demonstrate both types of bars and each of the bars contain many possible types of regular orbits, there is always a dominant one that is responsible for particular morphological features. 
\par
One of the most important results of our research is that the dominance of one or another orbital group is determined, first of all, by the potential of the galaxy. This initial idea turned out to be very fruitful and led us to the following conclusions.
\begin{enumerate}
\item 
In galaxies with classical bulges, albeit of small mass, the family of box-shaped orbits decreases in number, and the structure formed by these orbits becomes shorter. At the same time, the x1 orbital family, which support a long and narrow bar, is growing in number.
\item
In potentials with a high matter concentration, two `non classic' types of orbits come to the fore. These orbits do not support the bar, but apparently branch off from the well-known x1 and x2 families.
\item
One of these orbital group (\blu in our notation) maintains the rounded shape in the central area of the bar, turning the part of it into a barlens. If before that there were only qualitative discussions about the possible types of orbits that inhabit the barlens, now both the structure itself and the orbital groups supporting it are directly highlighted in the $N$-body models.
\item
The motion of stars belonging to this orbital group is most affected by the bulge potential. This group \textcolor{black}{is formed in situ in the presence of the compact bulge.}
We believe that this orbital group is a key component of a barlens, without which it is impossible to obtain its characteristic roundish isophotes, which are observed in many galaxies with a barlens. In terms of orbits shape, such orbits are very simple, but they have not explicitly stood out in such studies.
\item
The orbit language, which made it possible to distinguish the barlens as a separate structure, can be useful in constructing photometric models of galaxies. A modelled density profile could be inserted into packages for the photometric decomposition of galaxies. Dividing the bar into separate orbital groups will help to further study the difference in the vertical structure of the barlens and the rest of the B/PS bulge.
And finally, the orbital composition of the barlens opens up great opportunities for creating observational kinematic tests and further study of galaxies with barlenses.
\item
We also proposed an explanation of the unusual morphology of some of the galaxies with barlenses. For example, traces of X-shaped structures observed in galaxies with barlenses in unsharp-masked images can indicate that the structure of a barlens is determined by the delicate balance between the populations of different orbital groups, in particular, by the role of the box-shaped part of the bar. This is also evidenced by the existence of galaxies with barlenses, which are not so much rounded as `square'-like isophotes. Perhaps an `extended' part of the bl$_\mathrm{o}$ orbits is responsible for such isophotes.
\item
Finally, although the existence of x2 orbits in the central areas of the bar has been known for a long time, we were able to completely isolate the structure supported by these orbits. The length of this structure in the direction of the bar minor axis is quite large, and it seems to form two protrusions against the background of an elongated bar. We also showed that the population of this family\footnote{Strictly speaking, we deal with quasi-periodic orbits trapped around stable x2 orbits.} in our models increases with the transition to galaxies with a classical bulge. From an observational point of view, this is not strange, since it is in early-type galaxies with classical bulges that double bars are often found \citep{deLorenzo_etal2019}. However, in \textcolor{black}{stellar} $N$-body models, such a tendency is apparently revealed for the first time.
\end{enumerate}

\section*{Data availability}
The data underlying this article will be shared on reasonable request to the corresponding author.

\section*{Acknowledgements}
The authors express gratitude for the grant of the Russian Foundation for Basic Researches number 19-02-00249. We are grateful to the anonymous reviewer for a careful reading of the manuscript and many valuable comments that contributed to a improvement of the scientific quality of the manuscript and a clearer presentation of our results.

\bibliographystyle{mnras}
\bibliography{article4}

\appendix

\section{Extracting elongated \blu and \blo orbits}
\label{sec:surgery}

\begin{figure}
  \centering
  \includegraphics{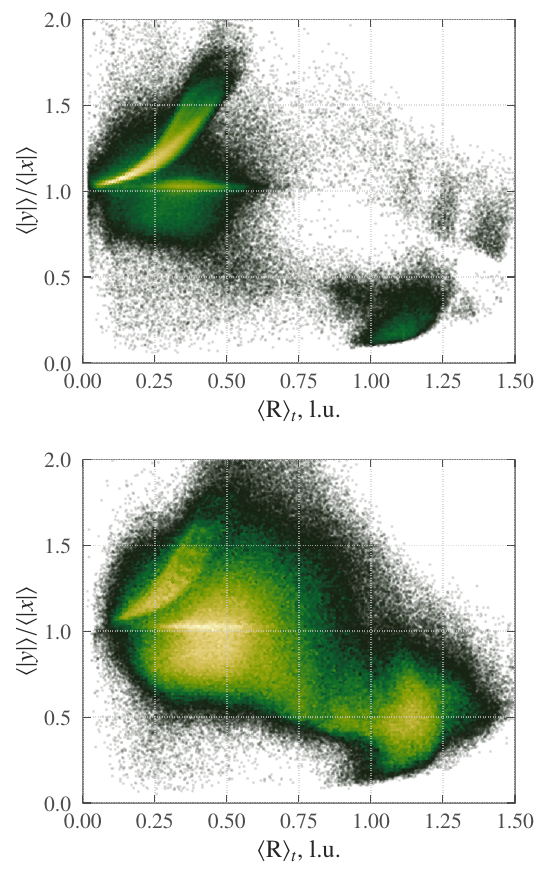}
  \caption{2D distribution of \blu (\emph{left panel}) and \blo (\emph{right panel}) orbits on 
    $\averg{R}_t - \averg{|y|}/\averg{|x|}$ plane for model BL calculated for the time interval $t = 400 - 500$.}%
  \label{fig:nonbar_surgery}
\end{figure}

When analysing the orbits supporting the barlens, we were faced with a situation where a small number of orbits elongated along the major axis of the bar were mixed with the main ensemble of orbits, which manifested itself as a roundish structure. Apparently, these elongated orbits are the orbits that have not yet had time to scatter into the bulge potential. They cannot be considered part of the barlens and thus should be excluded from \blu and \blo orbit groups. To exclude these orbits, we have used the 2D distribution of $\averg{R}_t$ and $\averg{|y|}/\averg{|x|}$. Such 2D maps are shown in Fig.~\ref{fig:nonbar_surgery} for \blu and \blo orbits.
One can see two ``islands'' on both panels of this plot, corresponding to the groups described above. 
The leftmost island is constituted by orbits that are close to the centre and contribute to the barlens, while the rightmost one corresponds to the elongated orbits described above.
For \blu orbits, the groups are well separated for all models, and finding an appropriate delimiting line is not a problem. However, for \blo orbits, these two parts are connected by an ``isthmus''. In this case, we draw the border approximately along the
middle of the isthmus. In all cases, we exclude the area below and to the right of the boundary, described by the linear equation
$\averg{|y|}/\averg{|x|} = \alpha \averg{R}_t + \beta$.

\label{lastpage}
\end{document}